\documentclass[aps,pra,twocolumn,showpacs,preprintnumbers,amsmath,footinbib,10pt]{revtex4-2}
\pdfoutput=1

\usepackage{graphicx}
\usepackage{dcolumn}
\usepackage{bm}
\usepackage[utf8]{inputenc}
\usepackage[T1]{fontenc}
\usepackage{mathptmx}
\usepackage{epsfig}
\usepackage{epstopdf}
\usepackage{amsmath}
\usepackage{txfonts}
\usepackage[sort&compress]{natbib}
\usepackage{rotating}
\usepackage{graphics}
\usepackage{array,multirow}
\usepackage[english]{babel}
\usepackage{attachfile}
\usepackage{pgfplots}
\usepackage[export]{adjustbox}
\usepackage{braket}
\usepackage{svrsymbols}
\usepackage[caption=false]{subfig}
\pgfplotsset{compat=1.15}
\hypersetup{urlcolor=purple, citecolor=purple, linkcolor=purple, colorlinks=true}

\def\be{\begin{equation}}
\def\ee{\end{equation}}
\def\bea{\begin{eqnarray}}
\def\eea{\end{eqnarray}}

\DeclareMathOperator\erfc{erfc}
\DeclareMathOperator\sinc{sinc}

\def\sinc{\text{sinc}} % <--- MathJax sinc definition on this line (see source code)
\def\brms{B_\text{rms}}
\def\frms{\phi_\text{rms}}

\begin{document}

\title{Correlated noise in Brownian motion allows for super resolution}

\author{Santiago Oviedo-Casado${}^{1,\atom}$,Amit Rotem${}^{1,\magnon}$,Ramil Nigmatullin${}^{2}$,Javier Prior${}^{3,4}$,Alex Retzker${}^{1}$}
\affiliation{${}^{1}$Racah Institute of Physics, The Hebrew University of Jerusalem, Jerusalem, 91904, Givat Ram, Israel\\
${}^{2}$Complex Systems Research Group and Centre for Complex Systems, Faculty of Engineering and IT, The University of Sydney, Sydney, NSW 2006, Australia.\\
${}^{3}$Departamento de F{\'i}sica Aplicada, Universidad Polit{\'e}cnica de Cartagena, Cartagena 30202 Spain\\
${}^{4}$Instituto Carlos I de F{\'i}sica Te{\'o}rica y Computacional, Universidad de Granada, Granada 18071, Spain}%
\email{${}^{\atom}$oviedo.cs@mail.huji.ac.il\\
${}^{\magnon}$amit.rotem1@mail.huji.ac.il}
%${}^{{\text{\Hygiea}}}$juergen.hauer@tum.de\\
%${}^{{\text{\Psyche}}}$javier.prior@upct.es}

\begin{abstract}
Diffusion broadening of spectral lines is the main limitation to frequency resolution in non-polarized
liquid state nano-NMR. This problem arises from the limited amount of information that can be
extracted from the signal before losing coherence. For liquid state NMR as with most generic sensing
experiments, the signal is thought to decay exponentially, severely limiting resolution. However,
there is theoretical evidence that predicts a power law decay of the signal’s correlations due to
diffusion noise in the non-polarized nano-NMR scenario. In this work we show that in the NV based
nano-NMR setup such diffusion noise results in high spectral resolution
\end{abstract}

\maketitle

\section*{Introduction}

Spectral analysis is of utmost importance in a wide variety of fields, from material science to biology and medicine.
Among the most widespread techniques to obtain structural information in the form of a spectrum is Nuclear Magnetic Resonance (NMR), which is nonetheless hindered by low sensitivity.
One promising approach to improve the capacities of NMR is to reduce the sample to the nano-scale. This technique, however, is still limited by the finite resolution of spectral features.
A possible solution is to use polarized samples as in conventional NMR \cite{Glenn2018,Bucher2018}, but this approach requires either large samples or a substantial increase in experimental complexity.
In this work we challenge the claim that working with nano-sized samples limits resolution, and provide analytical and numerical evidence supporting the viability of the non-polarized setup as an alternative route to nano-NMR.

NV centers have been used extensively in the past as quantum sensors for the implementation of the nano-NMR scheme \cite{Mamin2013,Laraoui2013,Muller2014,Ajoi2015,Lovchinsky2016,Boss2017,Schmitt2017,Pfender2017,Glenn2018}.
In particular, the use of quantum heterodyne (Qdyne) measurement techniques (know as well as synchronized measurements), together with a suitable data-analysis algorithm has demonstrated that resolving two close frequencies requires no more than accumulating a sufficient number of measurements \cite{Retzker2019,Gefen2019}.
These techniques, however, are computationally heavy since they need to solve a global maximization problem in a large dimensional space that grows linearly with the measurement time.

\begin{figure}
	\hfill
    \subfloat[\label{problem illustration:Lorenzian}]{\includegraphics[width=0.48\linewidth]{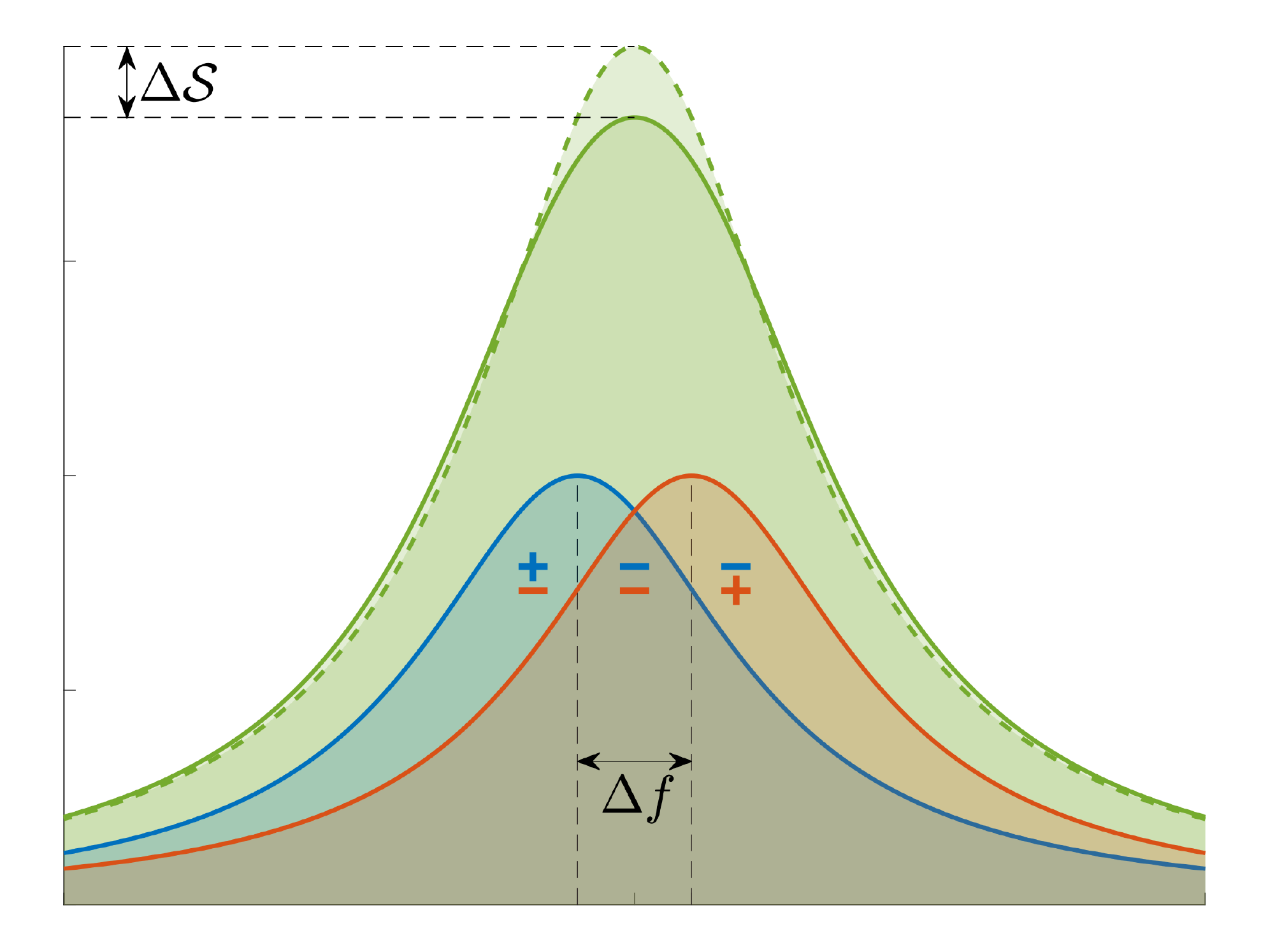}}
	\hfill
    \subfloat[\label{problem illustration:Daniel}]{\includegraphics[width=0.48\linewidth]{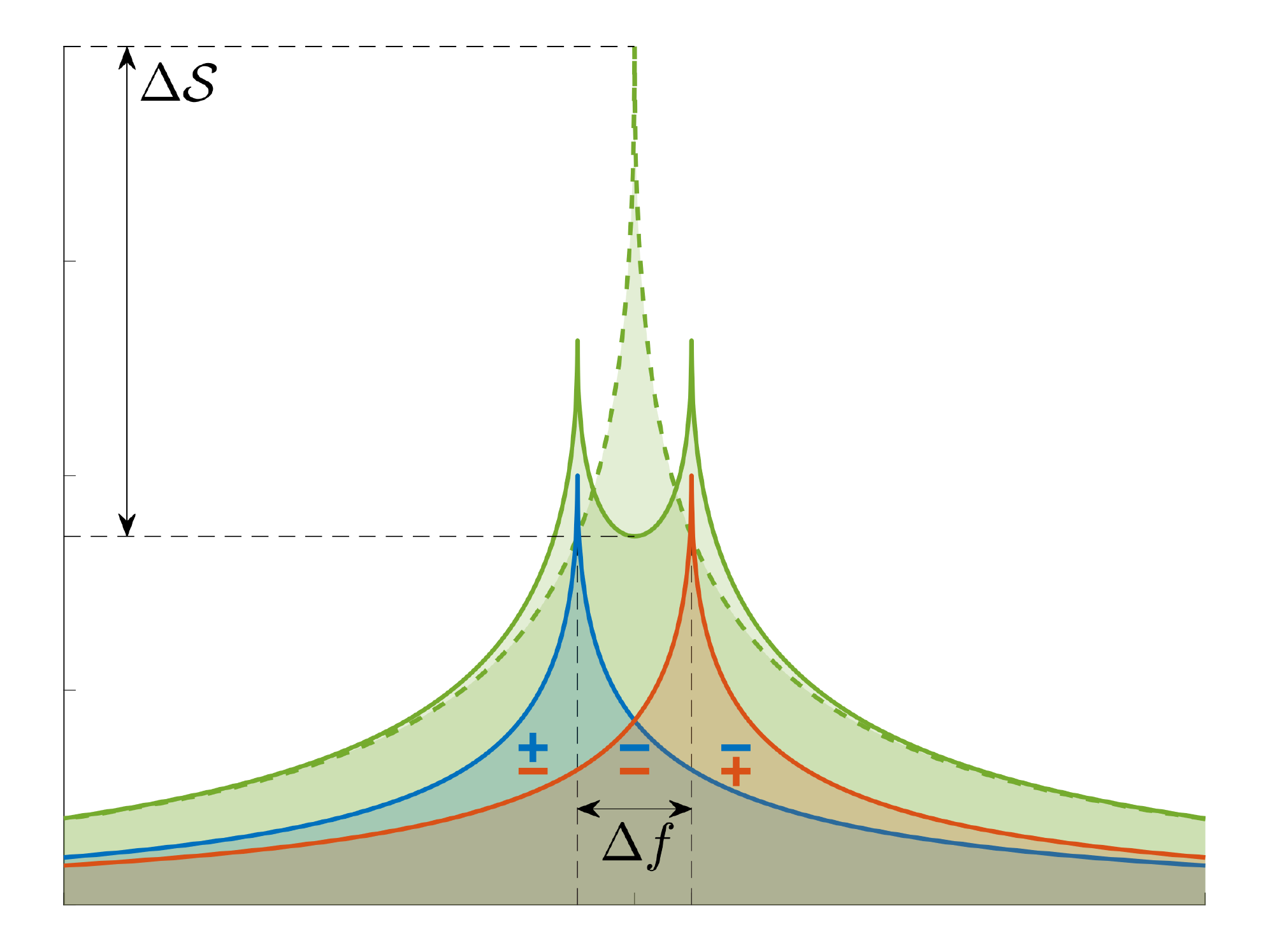}}
	\hfill
    \caption{
    Problem illustration. When the line-shapes of two underlying frequencies (blue and orange) overlap, the measured line-shape (solid green) can be very similar to line-shape of a single, strong, frequency (dashed green). The difference between the two line-shapes is most notable at the peak/center of the spectrum, where the changes brought about by the two underlying line-shapes coincide (blue and orange ``minus'' signs, indicating that \(\Delta \mathcal{S}\) is negative for a finite \(\Delta f\)), whereas at the edge of the spectrum the changes are opposite (blue ``plus'' and orange ``minus'' signs on the left, and vice versa on the right).
    (a) For a smooth function; e.g., Gaussian or Lorentzian, \(\Delta \mathcal{S} / \Delta f\)  is linear in \(\Delta f\) and thus small. In contrast, for a sharp-peak function as in (b)  \(\Delta \mathcal{S} / \Delta f \sim \Delta f^{-1/2}\), as can be shown from the diffusion dominated correlation function (Eq.~\ref{corr1}), and 
    resolution is not limited. See Appendix \ref{SM Problem illustration} for more details.
    }\label{problem illustration}
\end{figure}

%Measuring a spectrum that contains two (or more) similar frequencies that are closer than the characteristic width of their line-shape, known as Rayleigh Limit, results in a resolution problem (Fig.~\ref{problem illustration}).
Measuring a spectrum that contains two (or more) similar frequencies that are closer than the characteristic width of their line-shape results in a resolution problem (Fig.~\ref{problem illustration}).
The intuition behind the limited resolution can be understood in terms of the Rayleigh criterion from optics, where two images are resolvable only up to the wavelength used to image them. Here, the width of the line-shape plays the role of the wavelength.
This resolution problem for two close frequencies can best be understood by looking at the change in the spectrum (\(\mathcal{S}\)) as a function of the frequency difference. For a smooth function; e.g., a Lorentzian, a finite frequency difference has a very small effect on the spectrum (Fig.~\ref{problem illustration:Lorenzian}), whereas for a sharp-peak function the change is more pronounced (Fig.~\ref{problem illustration:Daniel}).
This suggests that for a sharp-peaked spectrum, spectral-resolution could be improved.

Spectral resolution in NV based liquid-state nano-NMR is limited mainly by the diffusion of nuclei in the sample \cite{Staudacher2013,Jelezko2015,Walsworth2016,Wrachtrup2017}.
When measuring a noisy signal oscillating at frequency \(\delta\), the amount of information that can be extracted from the auto-correlation of the signal; e.g., \(\cos(\delta t)C(t)\), is limited by the noise coherence time.
For diffusion noise in liquid state nano-NMR, \(C(t)\) is generally considered to be an exponentially decaying function leading to Lorentzian spectral line-shapes, impeding high spectral resolution. 
In this manuscript, we challenge this framework by building on the work of Cohen et al. \cite{Cohen2019}, which reported that a significant deviation from the Lorentzian line-shape paradigm occurs when measuring a magnetic field of a non-polarized nano-sized liquid sample with a shallow NV.
We show that diffusion does not limit resolution and that the analysis is computationally amenable and can be done with simple algorithms such as Fourier spectrum analysis.

The effect in \cite{Cohen2019} can be understood as follows.
The effective sensitivity of an NV located at depth \(d\) beneath a sample extends to a semi-sphere of radius \(d\) above the surface that contains \(N\propto d^3\) non-polarized nuclei.
The rms of the magnetic field sensed by the NV is thus \(\brms \propto \sqrt{N}/d^3\), where the \(d^3\) is due to the dipole-dipole interaction between NV and nuclei.
The peak of the power spectrum is thus \(S(\delta = 0) \propto \brms^2 T_\phi \propto 1/d\), with \(T_\phi \propto d^2\) the characteristic time that it takes the nuclei to diffuse out of the semi-sphere (i.e., the inverse of the signal bandwidth).
When, for example, applying dynamical decoupling (DD) sequence with detuning \(\delta\) from the nuclei Larmor frequency, a new length scale is introduced, i.e., \(\ell=\sqrt{D/\delta}\);
this length scale can be understood as a cut-off for the interaction between NV and distant nuclei; fields coming from these nuclei are slow changing and thus attributed to low frequency.
Using the same reasoning as before, the power spectrum around the peak is \(S(\delta) \propto 1/d-\alpha/\ell = 1/d - \alpha\sqrt{\delta / D}\), where \(\alpha\) is a positive number.
Therefore the power spectrum in NV based nano-NMR of liquid samples is a sharp-peaked function.
A similar effect has also been observed in diffusing atom systems \cite{Davidson2014}.
Conversely, in the time domain, where the resolution problem is manifested by our ability to see a beat-note, the measurement protocol with a shallow NV produces a correlation function with polynomial rather than exponential decay, such that the beating between close frequencies can be observed, allowing higher resolution.

\section*{FI analysis}
We now analyze the effect of long-lived correlations on frequency estimation and resolution. The resolution problem is characterized by an estimation error for the frequencies that diverges when the frequency difference is much smaller than the characteristic noise frequency, $T_\phi^{-1}$, as demonstrated by a vanishing amount of information extracted from the signal \cite{Tsang2016}.
For a noise that is a stationary Gaussian process, with a covariance function of the form \(\text{Cov}( t ) \propto C( t ) \sum_{j=1}^N \cos ( \delta_j t )\), the resolution problem occurs for \( | \delta_i -\delta_j | T_\phi < 1\).
We restrict the derivation to the estimation of a small single frequency $\delta$, which is a good model for the resolution problem since the average frequency is generally easier to estimate.
We analyze the three possible measurement scenarios, i.e. correlation spectroscopy \cite{Wrachtrup2015,Laraoui2013}, Qdyne/synchronized measurement protocol \cite{Schmitt2017,Glenn2018,Boss2017}, and power spectrum probing \cite{Romach2015}. For the full details of this derivation and schematics of each protocol we refer the reader to the Supplementary Information.

\subsection*{Correlation spectroscopy} The fluorescence response of the NV can be modeled by a Poisson distribution with a rate parameter that depends on the NV state (\(m=0,1\)).
In the correlation spectroscopy scenario, the average number of photons detected is given by \cite{Wrachtrup2015,Laraoui2013}
%For small measurement contrast we can approximate this response to a single Poisson distribution with varying rate parameter
\begin{equation}
	p = \eta + \frac{c}{2} \langle \sin (\phi_s ) \sin (\phi_{s+t} ) \rangle ,\label{correlation signal}
\end{equation}
where \(\eta,c\)  are the average detection rate and contrast, and \(\phi_s (\phi_{s+t})\) is the phase accumulated by the NV during the first (second) interrogation time (\(\tau\)). These phases are calculated by integrating over the magnetic field. We model the magnetic field as stationary Gaussian processes oscillating at frequency \(\delta\), with a characteristic correlation time \(T_\phi\) and a mean field strength of \(B_\text{rms}\). Averaging over realizations of the magnetic field yields
\begin{equation}
	p =  \eta + \frac{c}{2} e^{-\phi_{\text{rms}}^2} \sinh (\phi_\text{rms}^2 \cos ( \delta t ) C (t/T_\phi ) ), \label{averaged correlation signal}
\end{equation}
where \(C( \cdot )\) is the correlation function (envelope) of the phases.  The rms of the accumulated phase and its correlation function can be approximated by \(\phi_\text{rms} \approx \gamma B_\text{rms}\tau\) and \(C(t/T_\phi)\cos ( \delta t ) \approx \text{corr}(B_s,B_{s+t})\) for a short interrogation time \(\tau \ll T_\phi\), where \(\gamma\) is the gyromagnetic ratio of the NV. For a weak signal (i.e., \(\phi_\text{rms}^2 \ll 1\)) Eq.~\ref{averaged correlation signal} can be approximated by 
\begin{align}
	p \approx  \eta + \frac{c}{2} \phi_{\text{rms}}^2 \cos ( \delta t ) C( t/T_\phi).
\end{align}
The FI of \(\delta\) from a single measurement (a single choice of \(t\)) is given by 
\begin{equation}
	j_{\delta,\delta} \approx \frac{c^2}{4 \eta +c^2}\phi_{\text{rms}}^4 t^2 \sin^2 (\delta t ) C^2( t/T_\phi) , \label{FI from single measurement}
\end{equation}
in the weak signal regime.
Eq.~\ref{FI from single measurement} shows that the sine term is the reason for the limited resolution.
The maximum amount of information from a single measurement (for small \(\delta\)) depends on the correlation function.
An exponential decay imposes an optimal measurement time that scales as \(t^\text{opt}\propto T_\phi\); i.e., the longest time possible before the correlation is exponentially small.
Thus the information scales as \(j_{\delta,\delta} \propto \delta^2 T_\phi^4\), and vanishes for \(\delta \rightarrow 0\).
By contrast, for a slow polynomial decay (i.e., \(C(z) \propto z^{-n}\) for large \(z\) and \(0.5<n<1.5\), with $z$ henceforth being $z = t/T_{\phi}$) the optimal measurement time scales as \(t^\text{opt}\propto \delta^{-1}\); i.e., the correlations are significant enough such that the sine term poses no problems.
Thus the information scales as \(j_{\delta,\delta} \propto \delta^{2n-2}T_\phi^{2n}\), with a weaker dependence on frequency.
With respect to the measurement time, the information rate is \(j_{\delta,\delta}/T_\text{tot} \propto \delta^{2n-1}T_\phi^{2n}\); consequently, for correlations with \(n<1.5\) there is a slight improvement in resolution, and for \(n=1.5\), as in \cite{Cohen2019} (Eq.~\ref{corr1}), there is no improvement over exponential correlations.
For this reason it may be desirable to consider different measurement protocols.

\subsection*{Qdyne/Synchronized measurements} Further impro{\-}vement can be made considering a synchronized measurement protocol \cite{Schmitt2017,Glenn2018,Boss2017}. In this scenario, the fluorescence response of the NV has a detection rate of
\begin{equation}
	q_t = \eta + \frac{c}{2} \sin (\phi_t ). \label{Qdyne signal}
\end{equation}
Thus the average probability for measuring the pair \((y_s,y_{s+t} )\) of number of photons is 
\begin{equation}
	\langle q_s q_{s+t}\rangle = \eta^2 + \frac{c^2}{4} e^{-\phi_\text{rms}^2} \sinh \!\big(\phi_\text{rms}^2 \cos ( \delta t ) C (t/T_\phi ) \big). \label{Qdyne correlation signal}
\end{equation}
Estimating the signal using the covariance between the number of photons detected at different times, the information about \(\delta\) (from two measurements with a time difference \(t\)) is given by
%The FI of $\delta$, derived from this probability is given by
\begin{equation}
	j_{\delta,\delta} = \frac{c^4}{(4 \eta+c^2)^2} \frms^4 t^2 \sin^2 (\delta t ) C^2( t/T_\phi) + \mathcal{O}(\frms^6 ). \label{FI from pair of Qdyne measurement}
\end{equation}
%\begin{equation}
%	j_{\delta,\delta} = \frac{c^4}{16 \eta^2}\phi_{\text{rms}}^4 t^2 \sin^2 (\delta t ) C^2( t/T_\phi) + \mathcal{O}(\phi_{\text{rms}}^6 ) + \mathcal{O}(c^6). \label{FI from pair of Qdyne measurement}
%\end{equation}

\begin{figure}
    \centering
    \includegraphics[width=\linewidth]{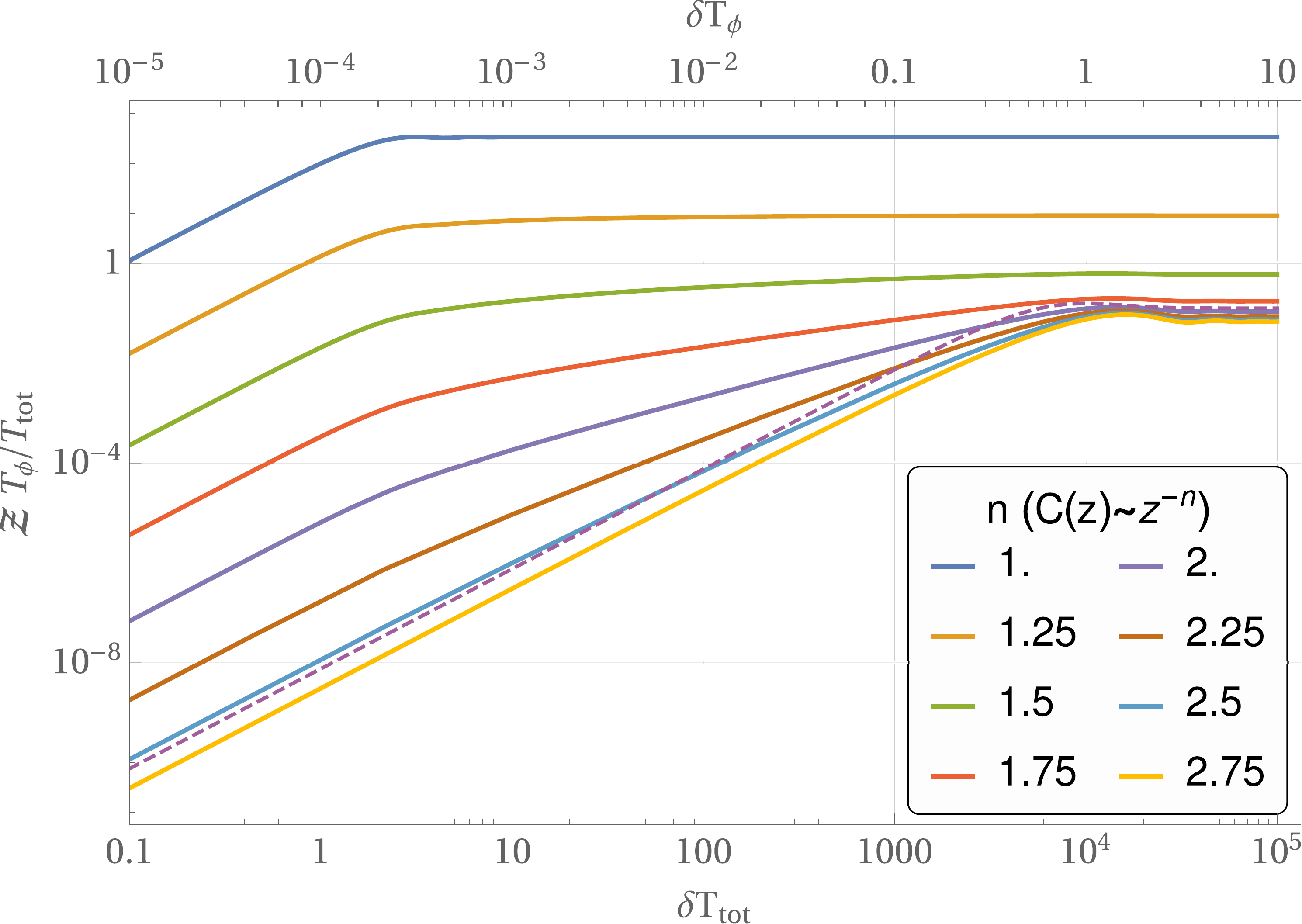}
    \caption{Scaling of the FI rate about \(\delta\) as a function of \(\delta\) (Eqs.~\ref{FI sum},\ref{z integral}); for this plot we set \(T_\text{tot}=10^4 T_\phi\). Different polynomial scalings are presented in different colors. The case of exponential correlation is presented as a dashed line. The information per unit of time saturates for \(\delta T_\text{tot} \gtrsim 1\), for correlations with slow polynomial decay (\(n<1.5\)). For faster decays (\(1.5<n<2.5\)) the characteristic time changes continuously towards \(\delta T_\phi \gtrsim 1\) (see top horizontal axis). For the limiting case of \(n=1.5\) the information rate changes its behavior for \(\delta T_\text{tot} \gtrsim 1\), but only saturates for \(\delta T_\phi \gtrsim 1\), which is attributed to the small logarithmic correction \(\log (\delta T_\phi)\).}
    %\colorbox{VioletRed}{(see SM for scaling of \(1/\Delta \left( \delta \right)=1/\left(J^{-1}\right)_{\delta,\delta}\))}}
    \label{Z integral scaling}
\end{figure}

This FI is obtained for a weak signal by
(least-squares) fitting of the correlation function.
%estimating the signal using the covariance of the number of photons detected, the information of \(\delta \) is
%\begin{equation}
%	j_{\delta,\delta} = \frac{c^4}{16 \eta^2}\phi_{\text{rms}}^4 t^2 \sin^2 (\delta t ) C^2( t/T_\phi) + \mathcal{O}(\phi_{\text{rms}}^6 ) + \mathcal{O}(c^6). \label{FI from pair of Qdyne measurement}
%\end{equation}
With each additional  measurement (performed at time \(t+\tilde{\tau}\)) we effectively obtain \(t/\tilde{\tau}\) additional "measurements" by correlating with all previous measurements. For small rms we can safely assume that the noise in the "measurements" is uncorrelated. For data taken at times \(t_m = m\tilde{\tau}\), the total FI is given by 
\begin{align}
	J_{\delta,\delta} &\approx \frac{c^4}{(4 \eta+c^2)^2}\phi_\text{rms}^4 \frac{T_\phi^4}{\tilde{\tau}^2} \mathcal{Z} \label{FI sum},\\
    \mathcal{Z} &= \intop_0^{T_\text{tot}/T_\phi} z^2 \sin^2 (\delta T_\phi z ) C^2(z) \left(\frac{T_\text{tot}}{T_\phi}-z \right)\text{d}z \label{z integral},
\end{align}
where we assumed \(\delta \tilde{\tau}\) and \(\tilde{\tau}/T_\phi\) to be small. The behavior of the integral in Eq.~\ref{z integral} for small \(\delta\) depends on the correlation function. For an exponential decay, \(\mathcal{Z}\propto \delta^2 T_\phi T_\text{tot} \) in the regime of \(\delta T_\phi \ll 1 \ll \delta T_\text{tot}\), whereas for polynomial decay 
\begin{equation}
\mathcal{Z}\propto 
\begin{cases}
( {T_\text{tot}}/{T_\phi})^{4-2n}&, n<1.5 \\
{\delta T_\text{tot}} (\delta T_\phi )^{2n-4}&, 1.5<n<2.5 \\
\delta^2 T_\phi T_\text{tot} &, n > 2.5
\end{cases}\label{eqcases}
\end{equation} 
in other words, there is a minute correction for small \(\delta\) when the polynomial decay is slower than \(2.5\). For decay rates slower than \(1.5\) the information is independent of \(\delta\), and the information rate increases with time (\(\propto T_\text{tot}^{3-2n}\)) (see Fig.~\ref{Z integral scaling}). In the limiting case of \(n=1.5\), \(\mathcal{Z} \propto \log ( \delta T_\text{tot} ) T_\text{tot}/T_\phi\) and the correction grows logarithmically when \(T_\text{tot}\) is large.

Compared to the correlation spectroscopy in Eq.~\ref{FI from single measurement}, the information from synchronized measurements in Eq.~\ref{FI from pair of Qdyne measurement} suffers from an extra \(c^2/(4\eta+c^2)\) factor (which is small in current experiments) due to correlations being obtained at post-processing rather than on the NV.
Nevertheless, this factor is compensated for by the fact that more statistics are gathered in Qdyne; i.e., roughly a factor of \((T_\text{max}/\tilde{\tau})^2\), assuming correlation spectroscopy measurements are performed using sequential correlation times up to time \(T_\text{max}\).
For exponential decays \(T_\text{max} \sim T_\phi\) and \(T_\text{max} \sim \delta^{-1}\) for slow polynomial decays, as seen in Eq.~\ref{FI from single measurement}.
These extra statistics compensates the logarithmic correction for small \(\delta\), meaning that the resolution with Qdyne is not limited by \(T_\phi^{-1}\).

Note that for correlation spectroscopy the shortest correlation time is limited by the DD sequence (which must be shorter than the coherence time of the signal), whereas for Qdyne is limited also by the readout/initialization time (\(\tilde{\tau}-\tau \approx 2.1 \mu s\), see for example \cite{Boss2017}); for exponential correlations this limits the Qdyne technique for samples with coherence time longer than the readout time. But for a slow polynomial decay this induces only a small constant factor on the information, as most of the information comes from long-time correlations.

\subsection*{Power spectrum measurements}
In the power spectrum measurement scenario, the interrogation time, \(\tau\), must be increased beyond the correlation time of the noise, which in most cases is impossible since the coherence time of the NV (\(T_2^{\text{NV}}\)) is too short.
The fluorescence response of the NV is given by 
\begin{equation}
\langle y_\omega \rangle = \eta - \frac{c}{2}  \exp (-\frac12 \gamma^2 B_\text{rms}^2T_\phi \tau \mathcal{S}_\tau(\omega ) ), \label{Spectrum measurement signal}
\end{equation}
where \(\mathcal{S}_\tau(\omega )\) is the unit-less (normalized by \(T_\phi \tau\)) power spectrum (convoluted with the filter function defined by the DD protocol).
The restriction on the interrogation time poses an extra limit on the field strength being probed \(\gamma^2 B_\text{rms}^2T_\phi \tau \lesssim  1\) (i.e., a large rms value will saturate the signal exponentially fast).
In addition, the inverse interrogation time sets the resolution for this measurement protocol; i.e., in order to resolve a frequency difference \(\delta\) we must set \(\tau > \delta^{-1}\).

\begin{figure*}
    \subfloat[\label{Fig3histograms:a}]{\includegraphics[width=5.8cm, height=4.455cm]{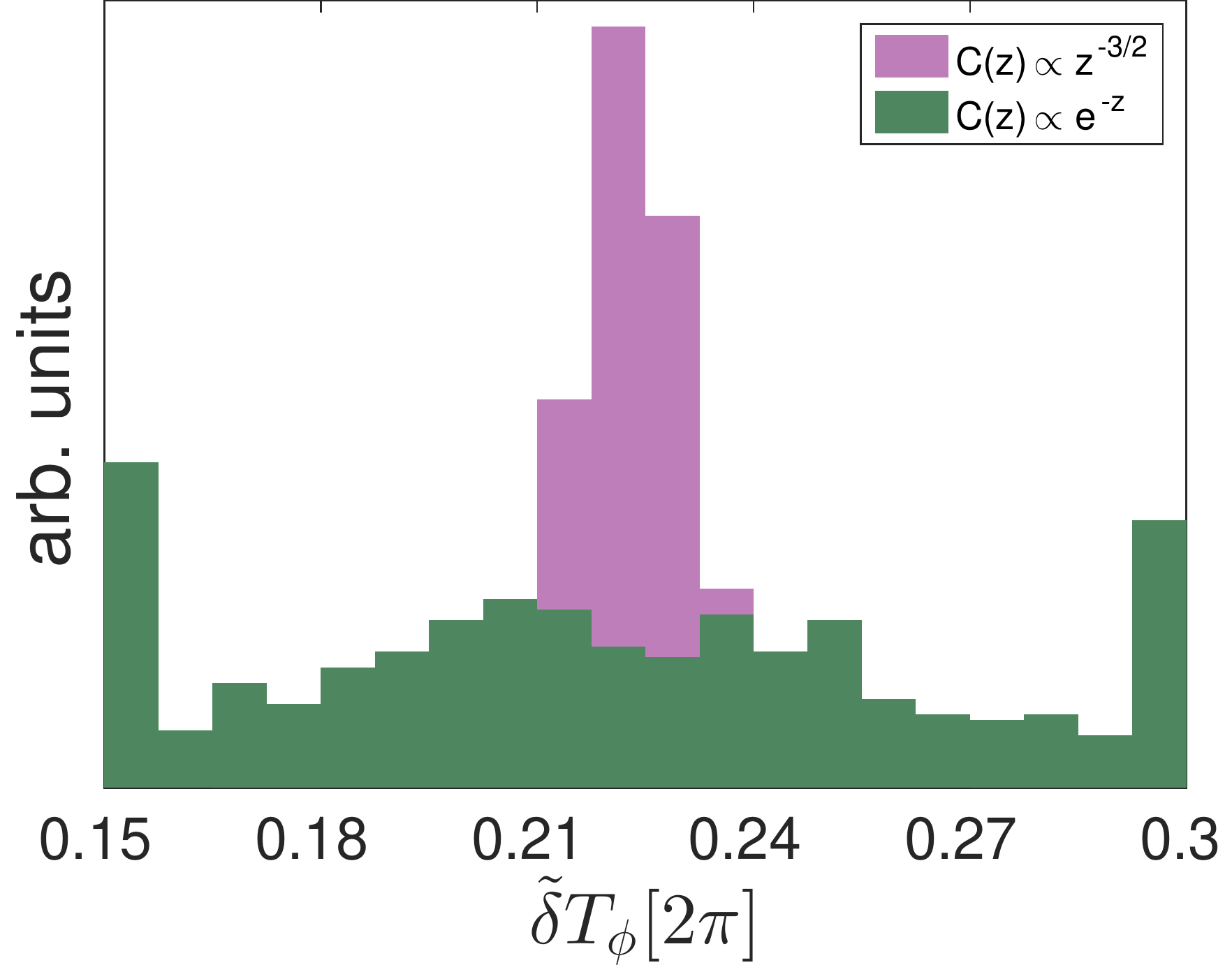}}
    \subfloat[\label{Fig3histograms:b}]{\includegraphics[width=5.8cm, height=4.455cm]{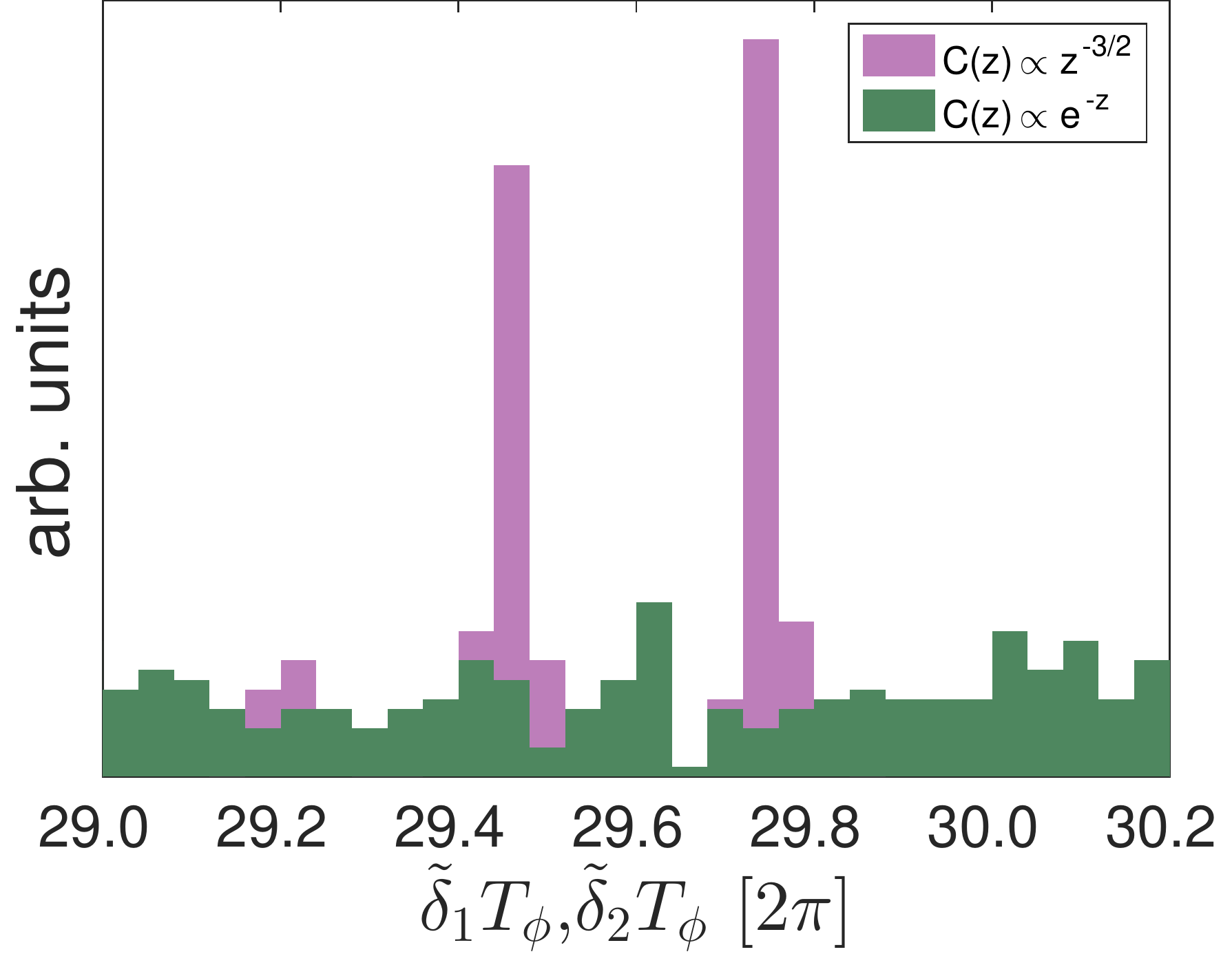}}
    \subfloat[\label{Fig3histograms:c}]{\includegraphics[width=5.8cm, height=4.455cm]{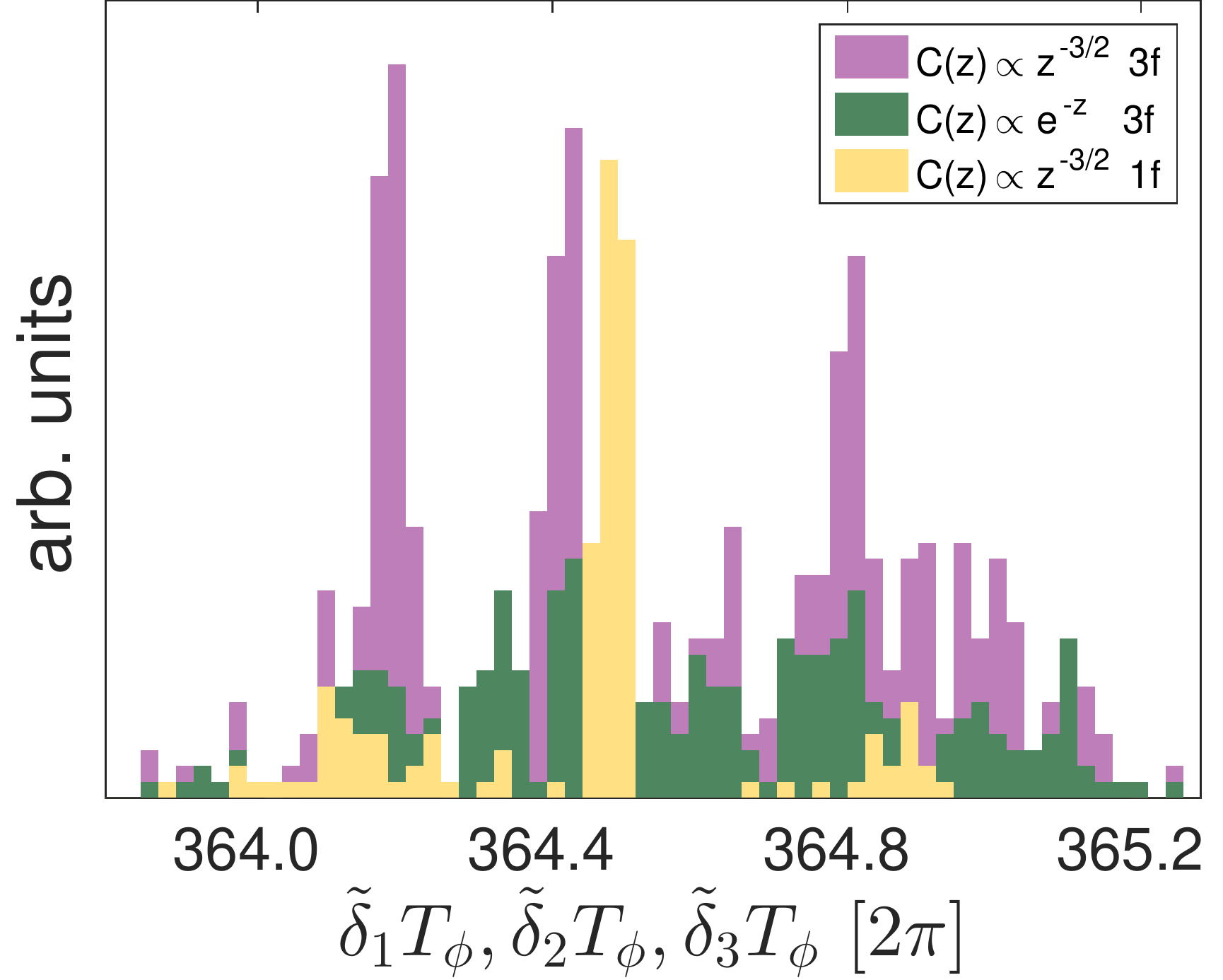}}
    \caption{
    (a) One frequency below the Rayleigh Limit is estimated for correlation \(C(z\gg1)\propto z^{-3/2}\) (purple) whereas estimation is not possible for exponential decay (green). $\phi_\text{rms}$ of the signal is 0.6. In purple, combinations of 50 estimation instances for each of the 12 different NV depths normalized to T$_\phi$. Signal noise in this case is generated by randomly taking vectors of length N from MD data (see Methods). In green, result for signals with the same parameters but with noise which is generated by fitting MD data to an exponential and fitting the signal to Eq.~\ref{corrfun1} with \(C(z\gg1)\propto z^{-3/2}\).
    (b) Two frequencies with a frequency difference ($\Delta\delta)T_\phi = 0.3 [2\pi]$ are resolved for long-lived correlations (purple) but remain unresolved for exponential decay (green). The amplitude of the signal is $\phi_\text{rms} \approx$ 0.6. Each histogram contains correlation function fittings of 200 measurement vectors with $2^{14}$ measurements.
    (c) Three frequencies (purple) with a frequency separation below the Rayleigh Limit, $(\Delta\delta) T_\phi \approx$ 0.3 [2$\pi$], are resolved for the case of long-lived correlations \(C(z\gg1)\propto z^{-3/2}\). For exponentially decaying correlations the same signal produces a histogram in which no single frequency can be pinpointed. In yellow, we generate a single-frequency signal. A signal with one frequency is estimated showing that the MSE is commensurate with the multi-frequency analysis.
    }\label{Fig3histograms}
\end{figure*}

When these requirements are met, the shape of the spectrum will dictate the information scaling; correlations that decay with a power law \(-n\) correspond to a spectrum that scales with a power law \(n-1\) around the peak. For a smooth spectrum (\(n>2\)) the information scales as the derivative of the spectrum (squared), \(j_{\delta,\delta} \propto T_\phi^2 (\delta T_\phi)^{\text{min}[2n-4,2]}\) at \(\omega=0\). For a sharp spectrum (derivative is discontinuous at the peaks, \(1<n<2\)) the optimal measurement is performed at \(\omega-\delta \propto \tau^{-1}\) (as close as possible to the peak, before the shape of the filter function starts to dominate) and the information scales as \(j_{\delta,\delta} \propto T_\phi^2 (\tau / T_\phi)^{4-2n}\). For the former case, resolution limit is set by \(T_\phi^{-1}\), albeit with a reduced "penalty", and by \(\tau^{-1}\) for the latter.

% \begin{enumerate}
%     \item For exponential correlations
%     \begin{enumerate}
%         \item[Corr:] \(J_{\delta T_\phi > 1} \lesssim \frac{c^2}{\eta} T_\phi T_{tot}\)
%         \item[Qdyne:] \(J_{\delta T_\phi > 1} \lesssim \frac{c^2}{\eta} \left(\frac{c^2}{\eta} \frac{T_\phi^2}{\tilde{\tau}^2} \right) T_\phi T_{tot}\)
%         \item[Spec:] \(J_{\delta T_\phi > 1} \lesssim \frac{c^2}{\eta} T_\phi T_{tot}\)
%     \end{enumerate}
%     \item For polynomial correlations
%     \begin{enumerate}
%         \item[Corr:] \(J_{\delta T_\phi > 1} \lesssim \frac{c^2}{\eta} T_\phi T_{tot}\)
%         \item[Qdyne:] \(J_{\delta T_{tot} > 1} \lesssim \frac{c^2}{\eta} \left(\frac{c^2}{\eta} \frac{T_\phi^2}{\tilde{\tau}^2} \right) T_\phi T_{tot} \log(\delta T_{tot})\)
%         \item[Spec:] \(J_{\delta \tau > 1} \lesssim \frac{c^2}{\eta} T_\phi T_{tot}\)
%     \end{enumerate}
% \end{enumerate}

%\input{C_corr_analysis}

\section*{nano-NMR signal analysis}

We now demonstrate resolution and verify the theoretical analysis by simulating and analyzing both single and multi-frequency signals. 
The procedure is as follows; first, we generate accumulated phases $\phi_t$ (Eq.~\ref{Qdyne signal}) by either using molecular dynamic (MD) simulations for a more  accurate description of an experimental situation (see Appendix \ref{MD}), or we sample a multivariate Gaussian distribution which simplifies the theoretical analysis. These phases are then used to simulate measurement vectors in a Qdyne protocol.
Parameter estimation is then performed by least squares fitting the signal correlation function to the theoretical model
\begin{equation}
	\sum_i (\phi_\text{rms}^{(i)})^2 \cos(\delta_i t + \varphi_i)C(t/T_\phi), \label{corrfun1}
\end{equation}
which corresponds to Eq.~\ref{Qdyne correlation signal} for weak signals. 
\(C( z )\) is considered either as polynomial correlations $\propto z^{-3/2}$ corresponding to Eq.~\ref{corr1} from \cite{Cohen2019} (henceforth \(C(z\gg1)\propto z^{-3/2}\)), or an exponential correlation \(\exp(-z)\) for comparison purposes. The $\varphi_i$ in Eq.~\ref{corrfun1} is a dummy parameter added for numerical reasons, and which tends to zero.
For more information about the numerical procedure see Appendix \ref{numerical calc}.

\subsection*{Resolution}Figure \ref{Fig3histograms} illustrates resolution beyond the Rayleigh Limit. We generate the signals of the magnetic field at different NV depths by using MD simulations of N $\approx$ 46k dipolar particles diffusing as a Lennard-Jones fluid, whose correlations behave as \(C(z\gg1)\propto z^{-3/2}\) at long times. Comparison to an exponential correlation function decay is done by fitting the MD results to an exponential model and using this model as a noise source.
In generating the signals, each NV-depth from MD is used, and is appropriately scaled according to the T$_{\phi}$ associated with the NV depth at which it is measured.
 Moreover, we work in the limit of $\delta T_\phi$ small ($\sim 0.3 [2 \pi]$) and small $\frms$ ($\sim$ 0.6), where as in the theoretical analysis shown in Eq.~\ref{z integral} the exponential correlations limit the resolution. 

\begin{figure*}[t!]
\centering
    \subfloat[\label{Fig4analysis:a}]{\includegraphics[width=0.48\linewidth]{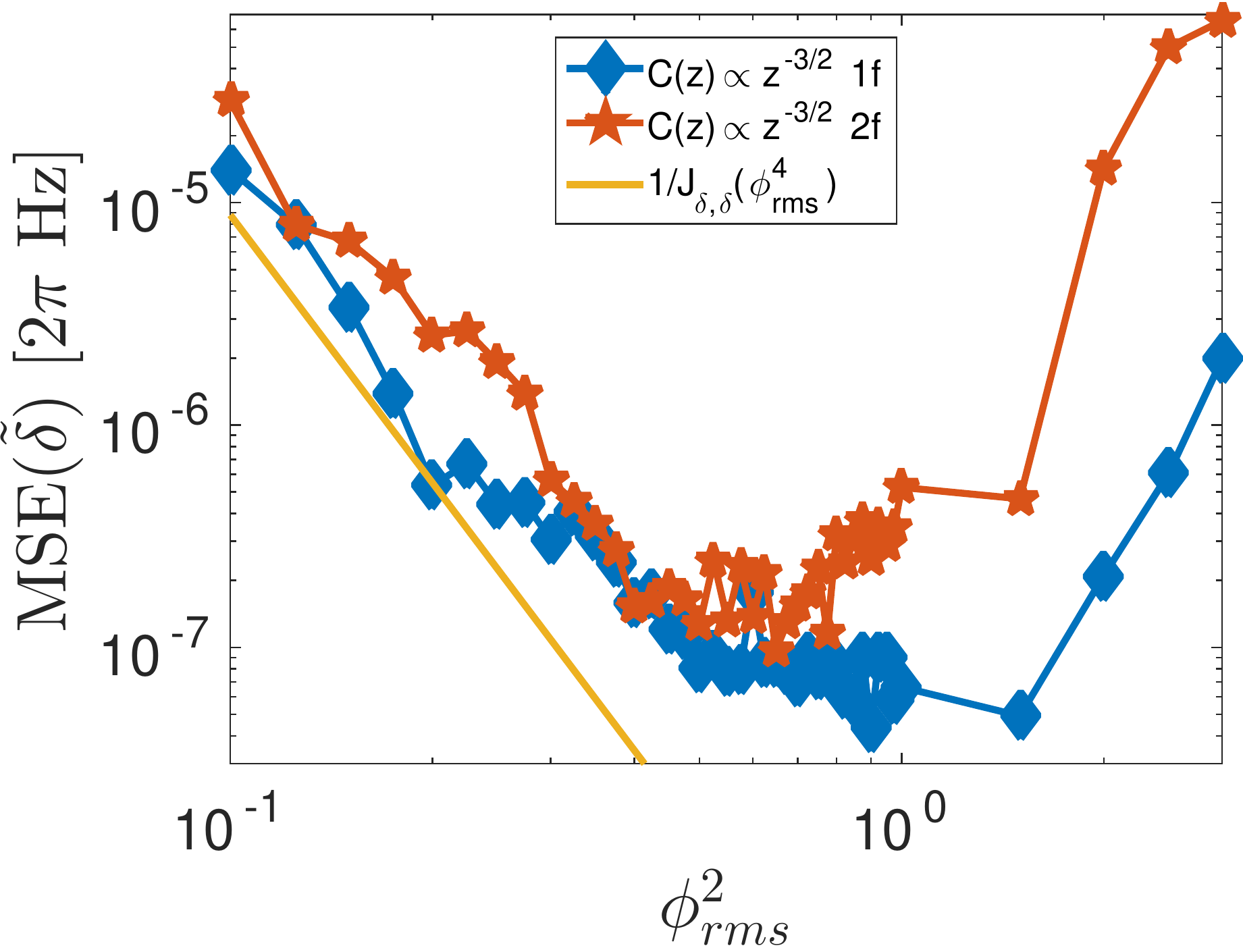}}
    \subfloat[\label{Fig4analysis:b}]{\includegraphics[width=0.48\linewidth]{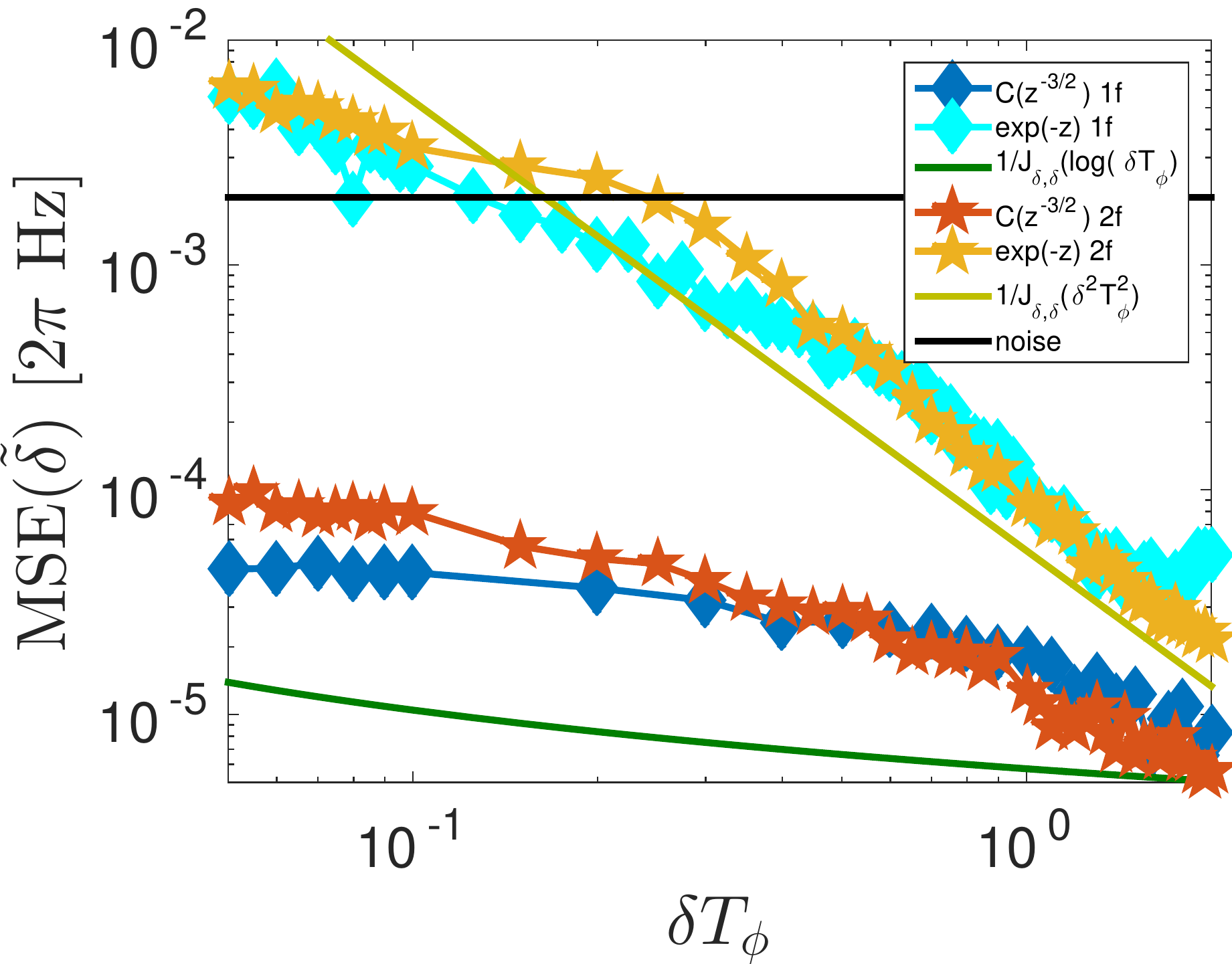}}
    \caption{(a) MSE of the frequency estimator $\tilde\delta$ (blue) and frequency difference estimator $\Delta\tilde\delta =|\tilde\delta_1 - \tilde\delta_2|$ (orange)  as a function of $\phi_\text{rms}$ with $\delta T_\phi$ = 0.5 [2$\pi$]. The line shows the theoretical prediction from Eq.~\ref{FI sum} (valid only for small $\frms$) dominated by $1/\phi_\text{rms}^4$. Below $\phi_\text{rms} \approx$ 0.1 the MSE saturates, indicating that the estimator is distributed across the whole search region. 
    (b) MSE of the frequency estimator $\tilde\delta$ and frequency difference estimator $\Delta\tilde\delta =|\tilde\delta_1 - \tilde\delta_2|$ for fixed $\phi_\text{rms}$ = 0.6 as a function of  $\delta T_\phi$ for polynomial \(C(z\gg1)\propto z^{-3/2}\) (diamonds) and exponential (stars) correlations. Horizontal line in (b) represents the flat histogram limit (noise level). Solid lines are the theoretical predictions from Eq.~\ref{eqcases} for n = 1.5 (dark green) proportional to $1/\log(\delta T_{tot})$, and exponential (light green) $\propto$ 1/$(\delta T_\phi)^2$.
    Each point represents the MSE of $2^8$ measurement vectors with $2^{14}$ measurements per vector. Note that the small differences between the one frequency and two frequency cases are merely numerical artifacts which would diminish for a higher number of measurement vectors. In both plots T$_{\text{tot}}/T_\phi \approx$ 164 for all points. In (a) $\delta T_\text{tot} \approx$ 82 [2$\pi$]. }
\label{Fig4analysis}
\end{figure*}

In Fig.~\ref{Fig3histograms:a} we depict the estimation of a single frequency for 600 measurement vectors, each composed of $2^{12}$ measurements. In fitting the correlation function Eq.~\ref{corrfun1}, a fitting is only accepted if $r^2 >$ 0.95. Fig.~\ref{Fig3histograms:b} depicts resolution for two close frequencies, which in this case loosely correspond to those of the experiment in \cite{Wrachtrup2017} but performed with an applied magnetic field one order of magnitude smaller. For this case we generate 200 measurement vectors of $2^{14}$ measurements each. A fitting is accepted if $r^2 >$ 0.95. In both cases, the frequencies were not resolved for the same parameters but rather with exponential correlations. 

%The complexity of parameter estimation is thought to increase exponentially with the number of frequencies.

Estimating close frequencies is a global optimization problem whose complexity increases exponentially in parallel with the size of the search space in which the frequencies live. In Fig.~\ref{Fig3histograms:c} we depict the resolution of three close frequencies which correspond to the frequencies from the experiment by Glenn et al. \cite{Glenn2018} but performed with a non-polarized sample. This is compared to a signal generated with exponential correlations, which does not allow for resolution of the frequencies. Furthermore, we include the histogram corresponding to a signal with one frequency slightly offset from the central frequency of \cite{Glenn2018}, generated with the same parameters and analyzed in the same way. It demonstrates that the Mean Square Error (MSE) is independent of the number of frequencies.
%Again the separation is $(\Delta\delta) T_\phi \approx$ 0.3 [2$\pi$] and $\phi_\text{rms} \approx$ 0.6.

\subsection*{Scaling analysis} We now proceed to the numerical analysis of the theoretical model presented in the previous section, in the case of one and two frequency signals. We show that for the anticipated signal in the nano-NMR scenario, the characteristic time for resolution is the total measurement time. In this case, we simulate synchronized measurements by generating signals with an analytical correlation function $C(t/T_\phi)$ where the noise comes from sampling a multivariate Gaussian distribution mimicking the scenario of small $\phi_\text{rms}$. We focus here on the case of n = 1.5 in Eq.~\ref{eqcases} corresponding to the correlation function in Eq.~\ref{corr1} (\(C(z\gg1)\propto z^{-3/2}\)) from \cite{Cohen2019}. A point in Fig.~\ref{Fig4analysis} corresponds to the MSE of a histogram composed of N = $2^8$ measurement vectors each, with $2^{14}$ measurements.

Figure \ref{Fig4analysis:a} displays the behavior of the MSE of the estimator as a function of $\phi_\text{rms}$. For fixed $\delta T_\phi= 0.5 [2\pi]$, below the Rayleigh Limit such that the signal with an exponential correlation could not be resolved, we simulate signals with varying $\phi_\text{rms}$.
According to Eq.~\ref{eqcases}, for a weak signal the MSE (i.e. 1/$J_{\delta,\delta}$) diverges as $\frms^{-4}$ as we observe in Fig.~\ref{Fig4analysis:b}, thus setting the optimal region for nano-NMR around $\phi_\text{rms}$ = 1. For strong signals, the information rate is exponentially suppressed. The scaling in the case of one frequency is not fundamentally different from that of two frequencies. 

In Fig.~\ref{Fig4analysis:b} we set $\phi_{\text{rms}} = 0.6$ and study the behavior with $\delta T_\phi$. Here we can observe the difference caused by extended correlations in the information rate and thus the resolution capacity. While for exponential correlations the MSE diverges quadratically with $\delta$, and rapidly saturates the histogram, for polynomial decays the divergence is slower. In the case of \(C(z\gg1)\propto z^{-3/2}\) the divergence is logarithmic in $\delta$ (see Eq.~\ref{eqcases}), as we see in Fig.~\ref{Fig4analysis:b}, i.e., it can easily be compensated for by increasing the measurement time. Note in addition that since $\frms$ ($\brms$) $\sim 1/d^{3/2}$ and $T_\phi \sim d^2$ \cite{Walsworth2016}, for \(C(z\gg1)\propto z^{-3/2}\) according to Eq.~\ref{eqcases}  the MSE is independent of the depth of the NV, as occurs with polarized nano-NMR.
%Together with $T_\phi$ scaling with depth ((Fig.~\ref{corr and parcorr:a} in \ref{noise}), this allows us to select the optimal NV for each experiment. 

\section*{Discussion}

We showed that spectral resolution in non-polarized liquid state nano-NMR is not necessarily limited by the broadening of spectral lines due to diffusion.
While for exponential correlations the resolution is limited by the inverse characteristic coherence time of the signal, we demonstrate that for (slow) polynomial correlations, as predicted by \cite{Cohen2019}, resolution is not limited. 

We analyzed the scenario in which the sensor is a shallow NV center.
In this case, the correlations decay as \(C(z\gg1)\propto z^{-3/2}\) at long times, producing sharp spectral features.
Moreover, increasing the number of frequencies analyzed does not hinder resolution.

Comparing the three measurement protocols we observe that for exponential correlations, the resolution problem always appears for \(\delta T_\phi < 1\), but the sensitivity of Qdyne is different by a factor of about \((c^2/\eta)(T_\phi/\tilde{\tau})^2\). For a low viscosity, water-like fluid this could still prove beneficial, despite the low contrast in state of the art systems (\(c^2/\eta \approx 0.016\)).
For power-law decay (with power of 3/2), while the sensitivity remains the same as the exponential case, the resolution capabilities of the power spectrum measurement and Qdyne protocols are extended. For power spectrum measurements, the protocol is limited by the time of a single measurement (\(\tau\)) which is only restricted by the coherence time $T_2$ of the NV sensor. The Qdyne protocol is virtually not limited by diffusion as the only limitation is the total measurement time.

The power law analysis presented here is so far based on theoretical grounds. Nonetheless, experimental evidence for a deviation from the exponential correlations paradigm already exist. In fact, Staudacher et al. found in \cite{Wrachtrup2015} a correlation function for a non-polarized liquid state nano-NMR experiment which exhibits a long-lived tail. Such behaviour was attributed to a surface effect which creates a thin layer of static, rotating molecules close to the surface of the diamond, finding a reasonably good agreement between the model and the experimental results. It is clear that the assumption of macroscopic Brownian motion with a Lorentzian profile and exponential correlations is too crude an approach to the non-polarized nano-NMR setting. As such, the diffusion induced long-lived correlations described in \cite{Cohen2019}, which we have demonstrated lead to enhanced resolution, are but a lower limit on the achievable resolution scaling of the non-polarized nano-NMR setup. Different physical effects such as those described in \cite{Wrachtrup2015} demonstrate that even longer-lived correlations can be expected to exist. As our analysis demonstrates, harnessing these power-law correlations leads to an increase of the information gathered (see Fig. \ref{Z integral scaling}), resulting in even better scaling for resolution of frequencies in a nano-NMR spectra.

\section*{Acknowledgements}

This project was supported by funding from the European Union Horizon 2020 Research and innovation Programme ERC grant QRES under grant agreement No 770929. and the collaborative European project ASTERIQS. S.O.C.is supported by the \textit{Fundaci{\'o}n Ram{\'o}n Areces}  postdoctoral fellowship (XXXI edition of grants for Postgraduate Studies in Life and Matter Sciences in Foreign Universities and Research Centers 2019/2020). J.P. is grateful for financial support from MCIU (SPAIN), including FEDER (Grant Nos. PGC2018-097328-B-100) together with Fundación Séneca (Murcia, Spain) (Project No. 19882/GERM/15).

\clearpage
%\newpage
%\input{S_supplemental-content}

\onecolumngrid
\appendix
\label{SM}

\section{Gaussian noise model}
We consider a two level system (TLS) coupled to a control field and a noisy field, in the interaction picture with respect to the TLS energy gap. We describe it as the Hamiltonian
\begin{equation}
	\mathcal{H}=\Omega(t) \sigma_\zeta + \frac{\gamma}{2} B (t)\sigma_z , \label{sigma_z Hamiltonian}
\end{equation}
where \(\Omega(t)\) represents the controls being applied to the sensor (\(\zeta\) is perpendicular to \(z\)). The second term is the noise signal that we want to measure, with \(\gamma\) the coupling constant of the field. We assume that the field \(B\) is a stationary Gaussian process with mean zero and covariance 
\begin{equation}
	\text{cov}(B_s,B_{s+t} ) = B_\text{rms}^2 \mathcal{C}(t) ,
    \label{B covariance}
\end{equation}
where \(B_\text{rms}^2\) is the variance of the field, and \(\mathcal{C}(\cdot)\) is the normalized covariance function.

Following the derivation by Cywi{\'n}sky et al. \cite{DasSarma2008}, we consider the control field as a sequence of (infinitely fast) \(n\) \(\pi\) pulses at times \(\{t_j\}_{j=1}^n\); e.g., for a CPMG sequence \(t_j=(j-\frac{n+1}{2}){\tau}/{n}\), about an axis that is perpendicular to \(z\). We define \(t_0=-\tau/2\) (\(t_{n+1}=\tau/2\)) as the start (end) of the sequence. The Hamiltonian in the interaction picture with respect to these pulses is given by
\begin{equation}
	\mathcal{H}_I = h(t) \frac{\gamma}{2} B (t)\sigma_z ,
\end{equation}
where \(h(t)\) is the response function, which for times \(t_j<t<t_{j+1}\) is equal to \(+1\) (\(-1\)) for even (odd) \(j\) and zero for \(t<t_0\) or \(t>\tau\); e.g., for CPMG \(h(t)=\theta(t+\tau/2)\theta(\tau/2-t)\text{sgn}(\cos((t-\tau/2) n \pi/\tau))\), where \(\theta(\cdot)\) is the Heaviside step function, and \(\text{sgn}(\cdot)\) is the sign function. The accumulated phase on the qubit
\begin{equation}
	\phi(t) = \gamma \intop_{-\tau/2}^{\tau/2} \text{d} a \, h(a) B (t+a)  ,
    \label{accumulated phase under DD}
\end{equation}
is a stationary Gaussian process with mean zero and covariance
\begin{align}
	\text{cov}(\phi_s,\phi_{s+t} ) &= \gamma^2 B_\text{rms}^2  \intop_{-\tau/2}^{\tau/2} \text{d} a \intop_{-\tau/2}^{\tau/2} \text{d} b \, \mathcal{C}(t+b-a) h(a)h(b) \label{phi covariance bis}\\
    &= \gamma^2 B_\text{rms}^2 \intop_{-\infty}^\infty \text{d}f \, \mathcal{S}(f) F(f) e^{i 2 \pi f t} ,
    \label{phi covariance}
\end{align}
where \(\mathcal{S}(f)=\intop_{-\infty }^\infty \mathcal{C}(t) e^{-i2\pi f t} \text{d}t\) is the power spectrum of \(B(t)\).  \(F(f)=\left|\widetilde{h}(f) \right|^2\) is the filter function which is defined by the pulse sequence \(\widetilde{h}(f)=\intop_{-\tau/2}^{\tau/2} h(t) e^{-i2\pi f t} \text{d}t\); e.g., for CPMG
\begin{align}
F(f) &=\frac{4}{\pi^{2} f^{2}}\frac{\sin^2 \!\!\left(\pi f \tau+\frac{n\pi}{2} \right) \sin^4 \!\!\left(\frac{\pi f \tau}{2n} \right)}{\cos^2 \!\!\left(\frac{\pi f \tau}{n}\right)} \\
&=\left| \frac{2\tau i^n}{\pi}\sum_{m=-\infty}^{\infty} \frac{(-1)^{(n+1)m}}{1+2m} \sinc \! \left(\left(f-(1+2m)\frac{n}{2 \tau} \right) \pi \tau \right)\right|^2 
\\ & \approx \frac{4 \tau^2}{\pi^2} \left(\sinc\! \left(\left( f-\frac{n}{2 \tau} \right) \pi \tau \right) +(-1)^{n}\sinc\! \left(\left( f+\frac{n}{2 \tau} \right) \pi \tau \right) \right)^2.
\end{align}
The main peaks of this function are located at \(f= \pm n/(2\tau)+\mathcal{O}(n^{-1})\), with a full width half max of about \(1/\tau\), and an area of \(4\tau / \pi^{2}\) each.

Eq.~\ref{phi covariance} has two regimes of interest to this manuscript. When the width of the filter function (\(\tau^{-1}\)) is smaller than the bandwidth of the signal (\(W\)), and when filter function is wider. They are denoted as 
\begin{equation}
	\text{cov}(\phi_s,\phi_{s+t} ) \propto \gamma^2 B_\text{rms}^2
	\begin{cases}
		 \tau \mathcal{S}(f_\text{DD})  &, \tau^{-1} \ll W \\
         \tau^2 \mathcal{C}(t) &,  \tau^{-1} \gg W
	\end{cases} 
\end{equation}
where \(f_\text{DD}=n/(2\tau)\) is the probing frequency (DD frequency). The former regime is appropriate for direct spectrum measurement, and the latter for correlation spectroscopy and synchronized measurement protocols. The following sections analyze the problem of resolution in these regimes.

\section{Resolution problem}

Resolution is defined as the ability to differentiate between close frequencies. To explore the resolution problem we focus on a simplified scenario where the signal (\(B(t)\)) is a narrow band noise and is composed of only two frequencies; i.e., \(B(t)=\sum_{k=1}^2 a_{k}(t) \cos (\omega_k t) + b_{k}(t) \sin (\omega_k t)\) where each \(\{a_k,b_k\}\) are stationary Gaussian processes with spectrum (\(S_k(f)\)) centered around \(f=0\). In this case the resolution problem emerges because of symmetries in the labeling (\(k\)) of the frequencies (i.e., \(1 \leftrightarrow 2\)) \cite{Retzker2019}, when the likelihood that the set of parameters \((\theta_1,\theta_2)\) that created the measurements set overlaps with the likelihood that the set \((\theta_2,\theta_1)\) created the same measurements. Here \(\theta_k\) denotes the set of parameters that characterize the process \(\{a_k,b_k\}\) (or equivalently the spectrum) and the frequency \(\omega_k\). We focus on the special case in which the processes \(\{a_k,b_k\}\) all have the same autocorrelation, and are all characterized by some coherence time \(T_\phi\) and signal strength \(B_\text{rms}\). Consequently the symmetries only affect the swapping of frequencies; i.e., \(\omega_1 \leftrightarrow \omega_2\). Generally speaking, the central frequency is easier to estimate \cite{Retzker2019}, so we reduce the problem further to that of estimating a single frequency (\(\delta\)) that is closer to zero as compared to the noise band-width (\(\approx T_\phi^{-1}\)); i.e., \(\delta \ll T_\phi^{-1}\).

We denote the general form of the signal considered in the rest of this manuscript
\begin{align}
	B(t) &= a(t) \cos (\delta t) + b(t) \sin (\delta t) \label{Bfield signle frequency} ,\\
    \text{cov}(B_s,B_{s+t}) &= \frac{4}{\pi^2} B_\text{rms}^2 \cos (\delta t) C(t/T_\phi) \label{cov single frequency},
\end{align}
and \(C(z)\) is either \(e^{-|z|}\) or
\[\begin{cases}
e^{-|z|} &, |z| \leq 1 \\
e^{-1} |z|^{-n} &, |z| > 1
\end{cases},\]
as an approximation for Eq. \ref{corr1} as calculated in \cite{Cohen2019}.

For short interrogation times \(\tau \ll T_\phi\) the response function simplifies to 
\begin{equation}
    	h(t)  =\theta(t+\tau/2)\theta(\tau/2-t) ,
\end{equation}
and the covariance Eq. \ref{phi covariance bis} can be written as
\begin{equation}
    \text{cov}(\phi_s,\phi_{s+t} ) = \tau^2 \sinc^2(\delta \tau/2) \text{cov}(B_s,B_{s+t} ).
\end{equation}

\section{Problem illustration in the spectrum \label{SM Problem illustration}}
The line shape for a polynomial correlation of power \((-n)\) with \(n < 3\), behaves as  \(S(f)\approx 1-\alpha |f|^{n-1}\) around the peak. For a noisy signal containing two frequencies the spectrum is given by \(\mathcal{S}\approx S(f+\Delta f) + S(f-\Delta f)\), and the derivative with respect to the frequency difference scales as \(\Delta f ^{n-2}\). On the opposite end, for a Gaussian or Lorentzian line shape we have that \(S(f)\approx 1-\alpha f^2\), and the derivative goes to zero as the frequencies overlap. This means that the diffusion process responsible for polynomial correlations as explained in Cohen et al. \cite{Cohen2019} does not limit the spectral resolution, as occurs in conventional NMR. In practice, other factors will limit the resolution, such as the measurement time (a single interrogation time) in power spectrum measurements. In what follows we analyze resolution in terms of these factors.

%For two underling frequencies the spectrum is given by \(\mathcal{S}\approx S(f+\Delta f) + S(f-\Delta f)\), and the derivative with respect to the frequency difference scales as \(\Delta f ^{n-2}\). For the case of Eq.~\ref{corr1}, \(n=3/2\), and this derivative diverges when \(\Delta f \rightarrow 0\). This is in contrast of a Gaussian or Lorentzian line shape, where \(S(f)\approx 1-\alpha f^2\), and the derivative goes to zero. This means that the diffusion process responsible for polynomial correlations as explained in the work of \citet{Cohen2019} does not limit the spectral resolution, as happens in conventional NMR. In practice, other factors will limit the resolution, such as the measurement time (a single interrogation time) in power spectrum measurements.

\section{Correlation spectroscopy \label{corr spectroscopy}}
We consider the following measurement protocol;
\begin{align}
	& \text{Initialize the NV to its ground state}, \nonumber\\
	& \text{pulses}:~\text{R}_y(\pi/2)-\text{DD}(f_\text{DD},\tau)-\text{R}_x(\pi/2), \nonumber\\
    & \text{wait}~(t-\tau), \nonumber\\
    & \text{pulses}:~\text{R}_y(\pi/2)-\text{DD}(f_\text{DD},\tau)-\text{R}_{-x}(\pi/2), \nonumber\\
    & \text{state readout},
\end{align}
where \(\text{R}_u(\theta)\) is a rotation of angle \(\theta\) around the \(u\) axis. \(\text{DD}\!\left(f_\text{DD},\tau\right)\) stands for some dynamical decoupling sequence at frequency \(f_\text{DD}\) with total duration \(\tau\). The dynamics during the DD sequences are given by the Hamiltonian in Eq.~\ref{sigma_z Hamiltonian}.
We assume that a \(T_2\) dephasing process erases the phase information during the wait time, but does not affect the state during the DD pulses sequence; meaning that the correlation time \(t\) is limited by \(T_1^{(NV)}\), and that the DD sequence time \(\tau\) is limited by \(T_2\).

\begin{figure}[!h]
   {\includegraphics[width=\columnwidth]{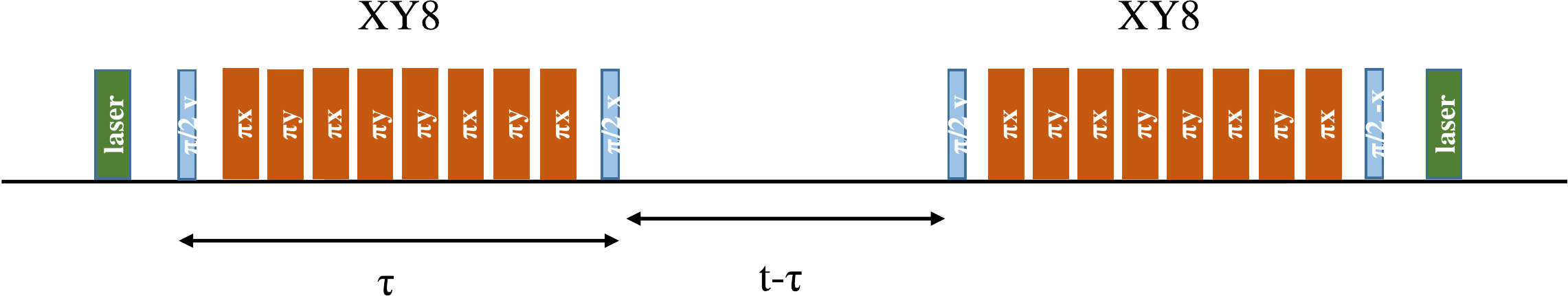}}
    \caption{Correlation spectroscopy measurement protocol. Following initialization of the NV via a 532 nm laser, a dynamical decoupling sequence gathers information about the sample and stores it the population of the NV. Following an erasure time, a second dynamical decoupling sequence gathers a second phase which then is correlated with the first one upon state readout.}\label{correlationspec}
\end{figure}

The probability of the NV to be in the excited state is
\begin{align}
	p_{s,t} &= \frac12 + \frac12 \sin (\phi_s ) \sin (\phi_{s+t} ),
\end{align}
where \(\phi_s \left(\phi_{s+t}\right)\) is the phase accumulated by the NV during the first (second) DD sequence (interrogation time) (Eq.~\ref{accumulated phase under DD}), and the time \(s\) represents some arbitrary initial time. We model the number of photons detected coming from the NV as a Poisson distribution with a rate that depends on the NV state
\begin{align}
	Y_{s,t} & \sim \text{Pois} ( \eta_{x_{s,t}} ) ,\\
    X_{s,t} & \sim \text{Bernoulli} ( p_{s,t} ),
\end{align}
where \(\eta_{0,1}\) is the average photon count from the NV \(m=0,1\) state. Given the stochastic nature of the phases (\(\phi_s\)) and the quantum nature of the system (\(x_{s,t}\)), the accessible distribution is the average photon count
\begin{align}
	P(y_t)&= \mathop{{}\mathbb{E}}_{x_{s,t},\phi_s,\phi_{s+t}}[P(y_{s,t})].
 %   \\ &= \frac12 \frac{\eta_1^{y_t} e^{-\eta_1}+\eta_0^{y_t} e^{-\eta_0}}{y_t !} \left(1 +\xi\frac{\eta_1^{y_t} e^{-\eta_1} - \eta_0^{y_t} e^{-\eta_0}}{\eta_1^{y_t} e^{-\eta_1}+\eta_0^{y_t} e^{-\eta_0}} \right)
\end{align}
 
\subsection{Estimation}
  The FI (sec.~\ref{FI defnition}) about the correlation function for correlation spectroscopy is given by 
 \begin{align}
 	J_{\mathcal{C}(t),\mathcal{C}(t)} &= \left( \frac{\partial \xi_t}{\partial \mathcal{C}(t)} \right)^2 \sum_{n=0}^\infty \frac{1}{2 n !} \frac{( e^{-\eta_0} \eta_0^n - e^{-\eta_1} \eta_1^n )^2}{e^{-\eta_0} \eta_0^n (1-\xi_t) + e^{-\eta_1} \eta_1^n (1+\xi_t)} , \label{SM - FI about corr spec} \\
    \xi_t &= e^{-\phi_{\text{rms}}^2} \sinh (\phi_{\text{rms}}^2 \mathcal{C}(t) ) , \label{xi definition}
 \end{align}
 where we denote \(\phi_\text{rms} \approx 2\gamma B_\text{rms} \tau / \pi\) for small \(\tau\). The sum in Eq.~\ref{SM - FI about corr spec} is bounded from above by \(((1-\xi_t)(\xi_t + \coth((\eta_0 + \eta_1)/2)))^{-1}\), saturating in a scenario with full measurement contrast (i.e., \(\eta_1 = 0\)  and \(\eta_0>0\)). For small measurement contrasts this sum is approximately \(c^2 /(4\eta)\).
 
 We use the sample mean to estimate the signal. The average photon count and variation are given by
\begin{align}
	\langle y_t \rangle &= \eta - \frac{c}{2} \xi_t, \\
    \text{Var}[y_t] &= \eta - \frac{c}{2} \xi_t + \frac{c^2}{4}(1-\xi_t^2), \\
\end{align}
where \(\eta = (\eta_0+\eta_1)/2\) is the average photon count, \(c=\eta_0-\eta_1\) is the contrast. Thus, the information about \(\mathcal{C}(t)\) (from the sample average of \(y\)) is given by 
\begin{align}
	J_{\mathcal{C}(t),\mathcal{C}(t)} &= \frac{1}{\text{Var}[y_t]}\frac{c^2}{4}\phi_\text{rms}^4 e^{-2\phi_\text{rms}^2} \cosh^2 (\phi_\text{rms}^2 \mathcal{C}(t)) \\ 
    & = \frac{c^2}{4\eta+c^2} \phi_\text{rms}^4 + \mathcal{O}(\phi_\text{rms}^6),
\end{align}
which coincides with the FI (Eq.~\ref{SM - FI about corr spec}) for small contrasts, and is relatively close (one order of magnitude) to the FI for large contrasts.

For the signal considered in this paper (Eq.~\ref{Bfield signle frequency}), the information on the frequency is given by
\begin{align}
	j_{\delta,\delta} &= J_{\mathcal{C}(t),\mathcal{C}(t)} t^2 \sin^2 (\delta t) C^2(t/T_\phi) \\
    & = \frac{c^2}{4\eta+c^2} \phi_\text{rms}^4 t^2 \sin^2 (\delta t) C^2(t/T_\phi) + \mathcal{O}(\phi_\text{rms}^6)
\end{align}

\section{Synchronized measurements \label{qdyne measurement}}

We consider a measurement protocol as follows;
\begin{align}
	& \text{Initializing the NV to its ground state}, \nonumber\\
	& \text{pulses}:~\text{R}_y(\pi/2)-\text{DD}(f_\text{DD},\tau)-\text{R}_{-x}(\pi/2), \nonumber\\
    & \text{state readout} + \text{clock readout}.
\end{align}
These measurements repeat in a synchronized fashion for each time \(\tilde{\tau}\), and \(\tau\) is the interrogation time. Accurately tracking the time between measurements enables us to correlate the measurement outcome in post-processing and estimate the signal.

\begin{figure}
   {\includegraphics[width=\columnwidth]{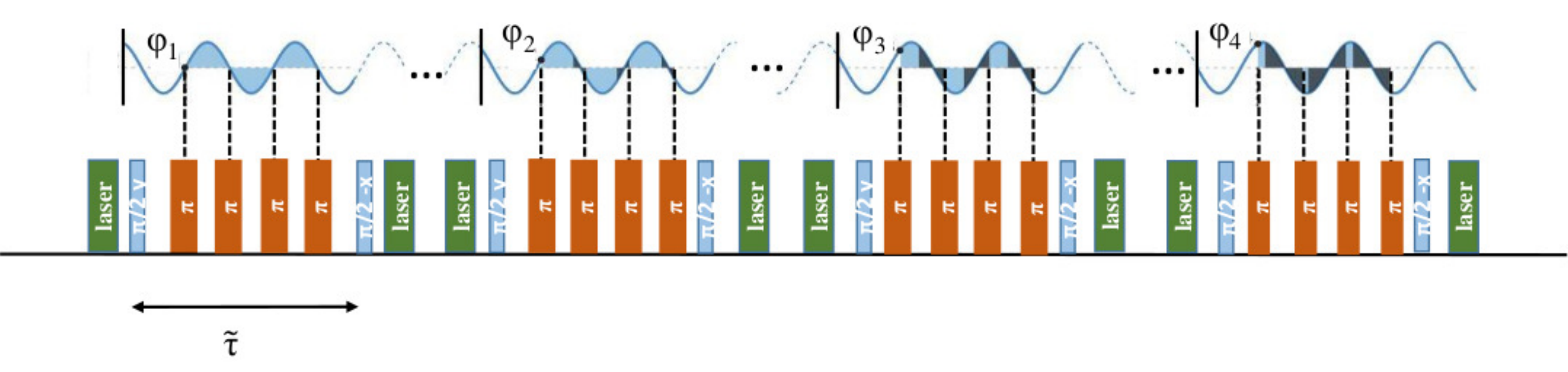}}
    \caption{Synchronize/Qdyne measurement protocol. A sequence of initialization - CPMG dynamcial decoupling - readout, is repeated to gather information about the correlations in the sample.}\label{qdyne}
\end{figure}

The probability of the NV to be in the excited state is
\begin{align}
	q_s &= \frac12 + \frac12 \sin (\phi_s)
\end{align}
where \(\phi_s\) is the phase accumulated by the NV during the DD sequence (interrogation time) (Eq.~\ref{accumulated phase under DD}) performed at time \(s\). We model the number of photons detected from the NV as a Poisson distribution with a rate that depends on the NV state
\begin{align}
	Y_s & \sim \text{Pois} ( \eta_{x_s} ) ,\\
    X_s & \sim \text{Bernoulli} ( q_s ),
\end{align}
where \(\eta_{0,1}\) is the average photon count from the NV \(m=0,1\) state. 

\subsection{Estimation}
%The average number of photons detected at each time is \(\eta\). 
In order to estimate the correlation function we use the covariance between consecutive measurements,
\begin{align}
	\text{cov} (n_s , n_{s+t}) &= \frac{c^2}{4} \xi_t  \\
    & = \frac{c^2}{4} \phi_{\text{rms}}^2 \mathcal{C}(t) + O(\phi_{\text{rms}}^4 ),
\end{align}
where \(\xi_t\) is as defined in Eq.~\ref{xi definition}. The sample covariance 
\begin{equation}
	S(t) = \frac{\tilde{\tau}}{T_\text{tot} -t} \sum_{s \in \{ \tilde{\tau},2\tilde{\tau},...T_\text{tot}-\tilde{\tau} \}} n_{s} n_{s-t} -\eta^2
\end{equation}
for different time differences (\(t\)) can no longer be considered an independent random variable, since it is calculated from a single time series \(\{ n_s \}\). The covariance between the sample covariance of different times is given by
\begin{align}
	\text{cov} (S(t_1) , S(t_2)) =& \mathop{{}\mathbb{E}}[S(t_1)S(t_2)] - \mathop{{}\mathbb{E}}[S(t_1)] \mathop{{}\mathbb{E}}[S(t_2)] \\
     =& \delta_{t_1,t_2}\frac{\tilde{\tau} \eta^{2}}{T_\text{tot}-t} {\bigg (}1+\left(\frac{c}{2\eta}\right)^{2}\left(\xi_{t_1}-\eta\xi_{0}\right) {\bigg )}+ \\ 
     & +\frac{\tilde{\tau} 2\eta}{T_\text{tot}-t} \left(\frac{c}{2}\right)^{2} (\xi_{t_1+t_2}+\xi_{t_2-t_1}) + \nonumber \\ 
     & -\frac{\tilde{\tau}}{T_\text{tot}-t}\left(\frac{c}{2}\right)^{4} {\bigg (} \xi_{t_1+t_2}+\xi_{t_2-t_1} + \nonumber \\
    & +\frac{1}{2}e^{-3\phi_{\text{rms}}^2} \cosh [2\phi_{\text{rms}}^2 (\mathcal{C}[t_1]+\mathcal{C}[t_2])] (e^{-\phi_{\text{rms}}^2 \mathcal{C}[t_1+t_2]}+e^{-\phi_{\text{rms}}^2 \mathcal{C}[t_2-t_1]})+ \nonumber \\
    & -\frac{1}{2}e^{-3\phi_{\text{rms}}^2} \cosh [2\phi_{\text{rms}}^2 (\mathcal{C}[t_2]-\mathcal{C}[t_1])] (e^{\phi_{\text{rms}}^2 \mathcal{C}[t_1+t_2]}+e^{\phi_{\text{rms}}^2 \mathcal{C}[t_2-t_1]}) {\bigg )} \nonumber \\
     = & \frac{\tilde{\tau} \eta^2}{T_\text{tot}-t} \left( \delta_{t_1,t_2} \left(1+ \frac{c^2 \phi_\text{rms}^2}{4 \eta^2} (\mathcal{C}[t_1] - \eta) \right) + \frac{c^2 \phi_\text{rms}^2}{2\eta} (\mathcal{C}[t_1+t_2]+\mathcal{C}[t_2-t_1]) + \mathcal{O}(\phi_\text{rms}^4) \right)
\end{align}
where \(t=\max (t_1,t_2)\), and the averaging is over the number of photons collected (\(n_s\)) over the distributions of \(x_{s,t},\phi_s,\phi_{s+t}\).

The information on \(\mathcal{C}(t)\) (from the sample covariance) is given by
\begin{equation}
	J_{\mathcal{C}(t),\mathcal{C}(t)} = \frac{1}{\text{cov} (S(t) , S(t))}\frac{c^4}{16} \left(\frac{\partial \xi_t}{\partial \mathcal{C}(t)} \right)^2.
\end{equation}

For the signal that is considered in this paper (Eq.~\ref{Bfield signle frequency}), the information on the frequency is given by
\begin{align}
	J_{\delta,\delta}  =& \frac{c^4}{16}\sum_{s,w} \frac{\partial \xi_w}{\partial \delta}  (([\text{cov} (S(t_1) , S(t_2))]_{t_1,t_2})^{-1})_{s,w} \frac{\partial \xi_s}{\partial \delta} \\
    = & \frac{c^4}{16 \eta^2} \sum_{t} \frac{T_\text{tot}-t}{\tilde{\tau}} \frac{\partial \xi_t}{\partial \delta} \frac{\partial \xi_t}{\partial \delta} + \mathcal{O}(\phi_\text{rms}^6).
\end{align}

\section{Power spectrum measurements}

In the scenario of power spectrum measurements we consider the measurement protocol;
\begin{align}
	& \text{Initializing the NV to its ground state}, \nonumber\\
	& \text{pulses}:~\text{R}_y(\pi/2)-\text{DD}(f_\text{DD},\tau)-\text{R}_{y}(\pi/2), \nonumber\\
    & \text{state readout},
\end{align}
such that the probability of the NV to be in the excited state is \(p = \cos^2(\phi/2) \); thus, the average photon detection rate is given by 
\begin{align}
	\langle y_\omega \rangle =& \eta - \frac{c}{2}  \exp (-\frac12 \phi_\text{rms}^2 \mathcal{S}_\tau(\omega ) ), \label{Spectrum measurement signal appendix} \\
  \text{Var}[y_\omega]  =& \langle y_\omega \rangle - \langle y_\omega -\eta\rangle^2 + \frac{c^2}{4},
\end{align}
where \(\phi_\text{rms}^2 = \gamma^2 B_\text{rms}^2T_\phi \tau\), and \(\mathcal{S}_\tau(\omega )\) is the unit-less (normalized by \(T_\phi \tau\)) spectrum convoluted with a filter function of width \(\tau^{-1}\).

\begin{figure}
   {\includegraphics[width=\columnwidth]{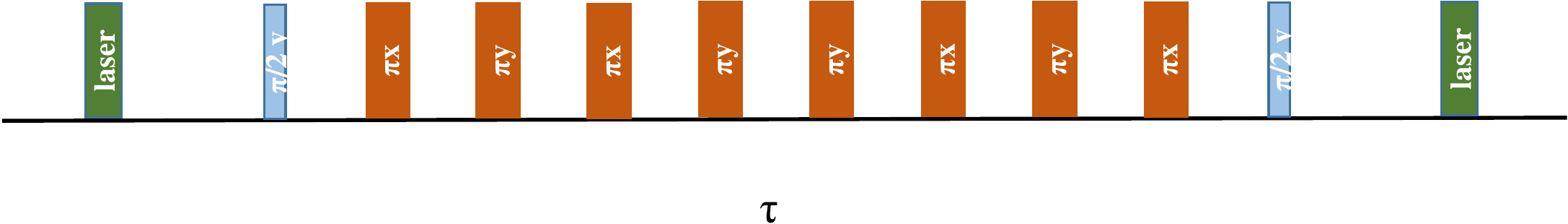}}
    \caption{Power spectrum measurements protocol following an XY8 dynamical decoupling sequence. A 532 nm laser initializes and reads out the NV state. The sequence duration $\tau$ is varyied to scan the power spectrum of the sample.}\label{powerspectrum}
\end{figure}

The measurement protocol in this scenario is similar to that of the correlation spectroscopy, with a different measurement basis and a longer interrogation time (\(\tau\gtrsim T_\phi\)), which could prove problematic if the coherence time of the sensor (\(T_2^\text{NV}\)) is short. Note that a strong field saturates the signal, which decays exponentially for large \(\phi_\text{rms}^2\).
The power spectrum can be approximated as \(\mathcal{S}_\tau( \omega )\propto 1-\alpha ( (\omega - \delta ) T_\phi )^{n-1} -\alpha ( (\omega + \delta ) T_\phi )^{n-1}\) for frequencies \(\tau^{-1}\lesssim | \omega\pm \delta | \lesssim T_\phi^{-1}\) and \(1<n<3\), where \(\omega\) is the detuning of the DD frequency from the central frequency in the spectrum.
For frequencies (\(\omega\)) closer to the peaks (\(\pm \delta\)), the spectrum behaves as the shape of the filter function, which is usually quadratic. At a distance \(T_\phi^{-1}\) from the peaks, the spectrum falls as \(\omega^{-2}\). For polynomial power \(n>3\) the spectrum behaves quadratically.
This means that the inverse interrogation time sets the resolution for this measurement, thus the interrogation time must be larger than \(\delta^{-1}\).

Under these restrictions (\(\gamma^2 B_\text{rms}^2T_\phi \tau \lesssim  1< \delta \tau\) and \(\tau \lesssim T_2^\text{NV}\)) , the information obtained about \(\delta\) using the average number of photons is
\begin{align}
j_{\delta,\delta} = \text{Var}[y_\omega]^{-1} (0.5 c \gamma^2 B_\text{rms}^2T_\phi \tau)^2 e^{-\gamma^2 B_\text{rms}^2T_\phi \tau \mathcal{S}_\tau(\omega )} \left(\frac{\partial \mathcal{S}_\tau( \omega )}{\partial \delta} \right)^2.
\end{align}
The (squared) derivative of the spectrum dictates how the information behaves.
When the spectrum is smooth (i.e., the derivative with respect to \(\omega\) is zero at the peak, \(n>2\)) the behavior of the (squared) derivative is similar to that of the Lorenzian case, but with a weaker dependence of \((\delta T_\phi)^{\text{min}[2n-4,2]}\) at \(\omega=0\).
When the spectrum is sharp (i.e., the derivative is discontinuous at the peak, \(1<n<2\)) the (squared) derivative scales as \((\tau/T_\phi)^{4-2n}\), at \(\omega=\delta-\tau^{-1}\).

For both the measurement resolution is set by \(\tau\) which is limited by \(\propto (\gamma^2 B_\text{rms}^2T_\phi)^{-1}\) or \(T_2^\text{NV}\).
An upper limit for the information is given when taking \((\gamma^2 B_\text{rms}^2T_\phi \tau)^2 e^{-\gamma^2 B_\text{rms}^2T_\phi \tau \mathcal{S}_\tau(\omega )}\approx1\),
\begin{equation}
j_{\delta,\delta} \lesssim \text{Var}[y_\omega]^{-1} c^2 T_\phi^2
\begin{cases}
(\delta T_\phi)^{\text{min[2n-4,2]}} &, n>2\\
(\tau/T_\phi)^{4-2n} &, 1<n<2
\end{cases}
\end{equation}
This means that for \(n<2\) the information is independent of \(\delta\) and the resolution is set by \(\tau^{-1}\), for \(n>2\) there is a reduced "penalty" for small \(\delta T_\phi\).

\section{Noise model for diffusing particles}\label{MD}

Each nucleus composing the sample substance interacts with the NV center via dipolar coupling; in the nano-NMR setting, nucleus dynamics manifests through the dephasing rate of the NV center. Calculating this dephasing rate involves solving the drift-diffusion dynamics equation. For an NV situated at a depth $d$ from the diamond surface and assuming that the liquid fills a semi-infinite volume above the diamond surface, the correlation function for the nucleus distribution is \cite{Cohen2019} 
%\begin{widetext}
\begin{equation}
%\begin{split}
C(z) = \frac{4}{\sqrt{\pi}} \Bigg( z^{-\frac{3}{2}} - \frac{3}{2}z^{-\frac{1}{2}} + \frac{\sqrt{\pi}}{4} + 3\sqrt{z} - \frac{3\sqrt\pi}{2}z + \sqrt{\frac{\pi}{z}}\erfc\Big(z^{-\frac{1}{2}}\Big)\exp{z^{-1}}\Big( - z^{-\frac{3}{2}} + z^{-\frac{1}{2}} - \frac{7}{4}\sqrt{z} + \frac{3}{2}z^{+\frac{3}{2}} \Big)  \Bigg),
\label{corr1}
%\end{split}
\end{equation}
%\end{widetext}
with $z = d^{-2} D t = {t}/{T_\phi}$, where $D$ is the diffusion coefficient for the fluid.

To accurately simulate the NV response signal to the magnetic field generated by a distribution of diffusing molecules used to demonstrate resolution in an experimental-like scenario, we perform molecular dynamics simulations.

For the molecular dynamics we consider $N\approx 46k$ dipolar particles within a simulation box of size $L_{x,y,z} \approx (50,50,24)$, with a NV located at depths in the range of $(0.3, 5)$. The particles within the box are simulated as a Lennard-Jones fluid with normalized parameters $\varepsilon=\sigma=1$, and are initialized into a thermal state at temperature T=1. During the simulation, the magnetic field induced by the particles at the NV position is measured along the $z$ direction for several NV depths.

% continue from here

\begin{figure*}
    \centering
    \subfloat[\label{corr and parcorr:a}]{\includegraphics[width=0.48\linewidth]{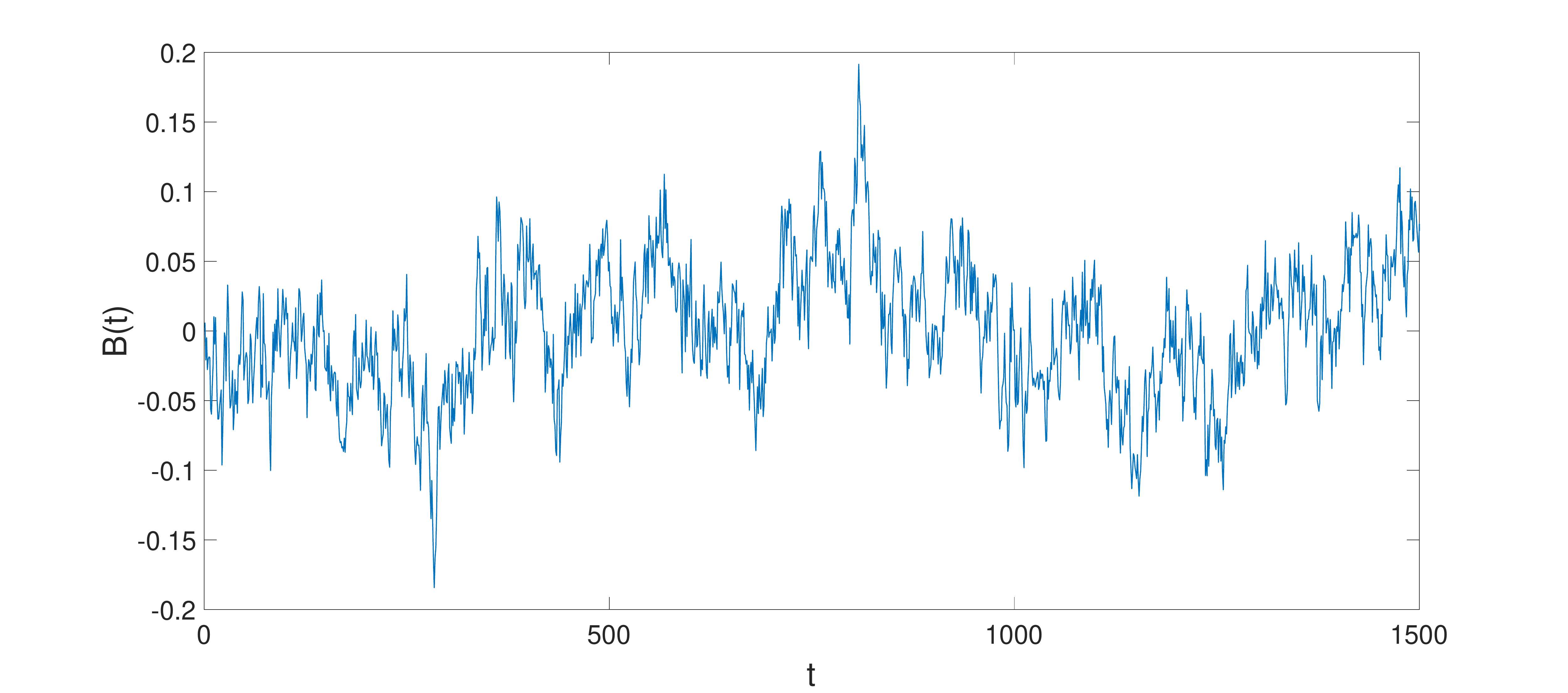}}\\
    \subfloat[\label{corr and parcorr:b}]{\includegraphics[width=0.48\linewidth]{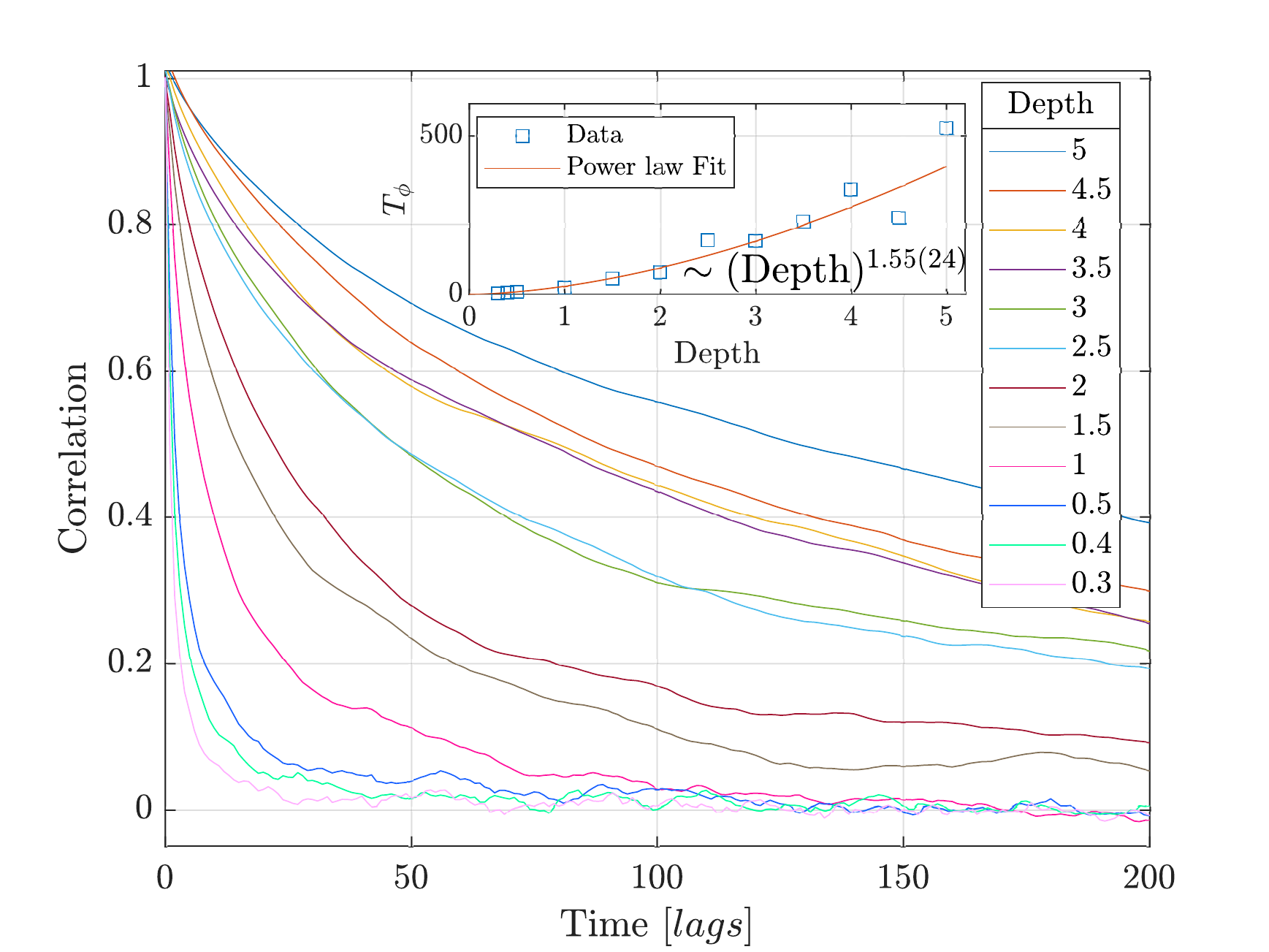}}
    \subfloat[\label{corr and parcorr:c}]{\includegraphics[width=0.48\linewidth]{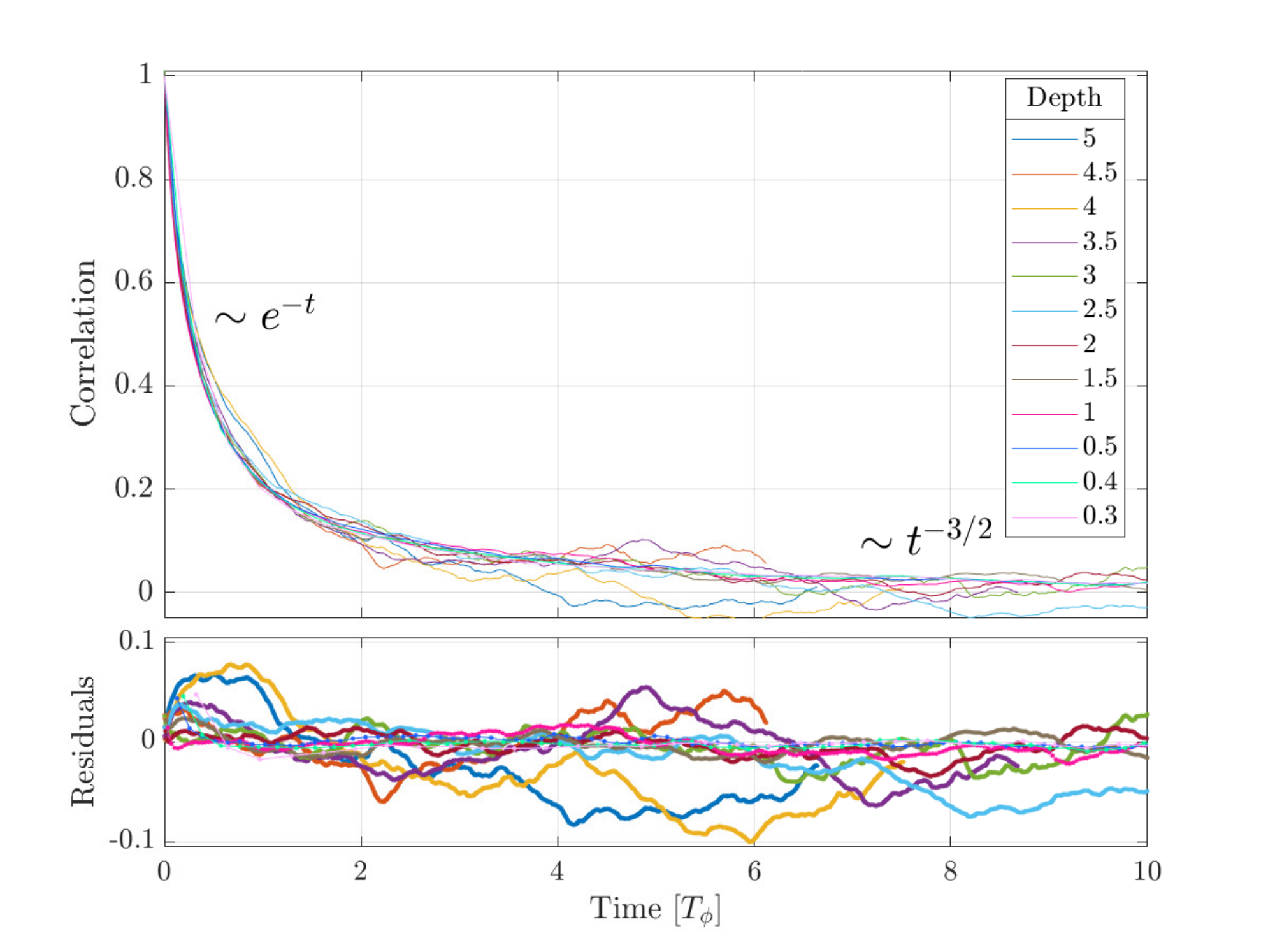}}
    \caption{(a) Sample of the molecular dynamics results for the magnetic field created at the NV position by a distribution of randomly diffusing dipolar particles.
    (b) Correlation function of the magnetic field created at the NV position for depths ranging from 0.3 to 5. Each correlation curve is the average of two realizations of molecular dynamics. The inset shows the correlation time of the magnetic field as a function of depth, obtained by fitting the correlation data to a correlation \(C(z\gg1)\propto z^{-3/2}\) (FT of Eq.~\ref{corr1}). Each $T_\phi$ is calculated as \(C(z = {t}/{T_\phi} \approx 0.255791) = {1}/{2}\).  Deviations from the theoretical exponent ($T_\phi \propto d^2$) occur due to finite box-size and simulation errors. Shallower NVs feature a different box-size; hence, a departure from a straight line.
    (c) Correlation of the magnetic field scaled to $T_\phi$. At short-times the correlation decays exponentially, whereas at long-times the decay is polynomial. This demonstrates that the diffusing particles create a highly correlated signal. Residuals are with respect to fitting in Fig.~\ref{corr and parcorr:a}. 
    }\label{corr and parcorr}
\end{figure*}

Analysis of the generated magnetic fields at different NV depths shows that the data have no trend and that the standard deviation remains scale invariant. This means we can compare different depths if appropriately scaled. This is done by calculating the correlations and partial correlations of the different time series. An example of a time series can be found in Fig. \ref{corr and parcorr:a}.

In Figure \ref{corr and parcorr:b} we analyze the temporal correlation in the magnetic field as a function of NV depth. The correlation, which is akin to the autocorrelation after correcting by the mean, tells us how a point in the time series is related to itself after k time-steps. We observe that it is highly dependent on the depth of the NV, as expected from the relation $T_\phi \propto d^2$ (the diffusion coefficient D is the same for all depths). Since resolution depends on $\delta T_\phi$ Fig. \ref{corr and parcorr:b} gives us information about which depth is more convenient, depending on the characteristics of the signal that we want to analyze. 

Fig. \ref{corr and parcorr:c} depicts the correlation corrected by depth. Note the deviation from exponential decay at long times, as described by Eq. \ref{corr1}, which is responsible for long-lived correlations. Moreover, this deviation is independent of the depth of the NV, which means that the same description is valid for the magnetic field at any NV depth.

When using MD vectors to simulate the accumulated phases $\phi_t$, we avoid correlations among different MD vectors by calculating each noise realization by randomly sampling two different instances of magnetic fields in the corresponding NV depth.

\section{Numerical calculations}\label{numerical calc}

Parameter estimation is done by numerically fitting each measurement vector to the theoretical model 
\begin{equation}
	\sum_i (\phi_\text{rms}^{(i)})^2 \cos(\delta_i t + \varphi_i)C(t/T_\phi)\label{eqG1}.
\end{equation}
The fitting is done by a non-linear least squares algorithm with finite-difference estimation of gradient. Each fitting is initialized with random values taken from uniform distributions around the mean signal values for each parameter in Eq.~\ref{corrfun1}. The width of the distributions coincides as well with the allowed search regions in the fitting process. These are, respectively, $\phi_{\text{rms}} \in$ [$\phi_{\text{rms}}^{\text{avg}}/2, 3\phi_{\text{rms}}^{\text{avg}}/2$], $\delta \in$ [$\delta^{\text{avg}}/2, 3\delta^{\text{avg}}/2$], $\varphi \in$ [0, 2$\pi$] and $T_\phi \in $[$T_\phi/100, 100T_\phi$]. Average values are estimated from the signal for the $\phi_{\text{rms}}$ or from the signal FT for $\delta$.

The $\varphi_i$ in Eq.~\ref{eqG1} is non-physical and is included for reasons of numerical stability. In all of the fittings it tends to either 0 or $2\pi$.

\section{Fourier Transform examples}\label{FT}

\begin{figure}
    \subfloat[\label{FigCFTfreqs:a}]{\includegraphics[width=5.8cm, height=4.455cm]{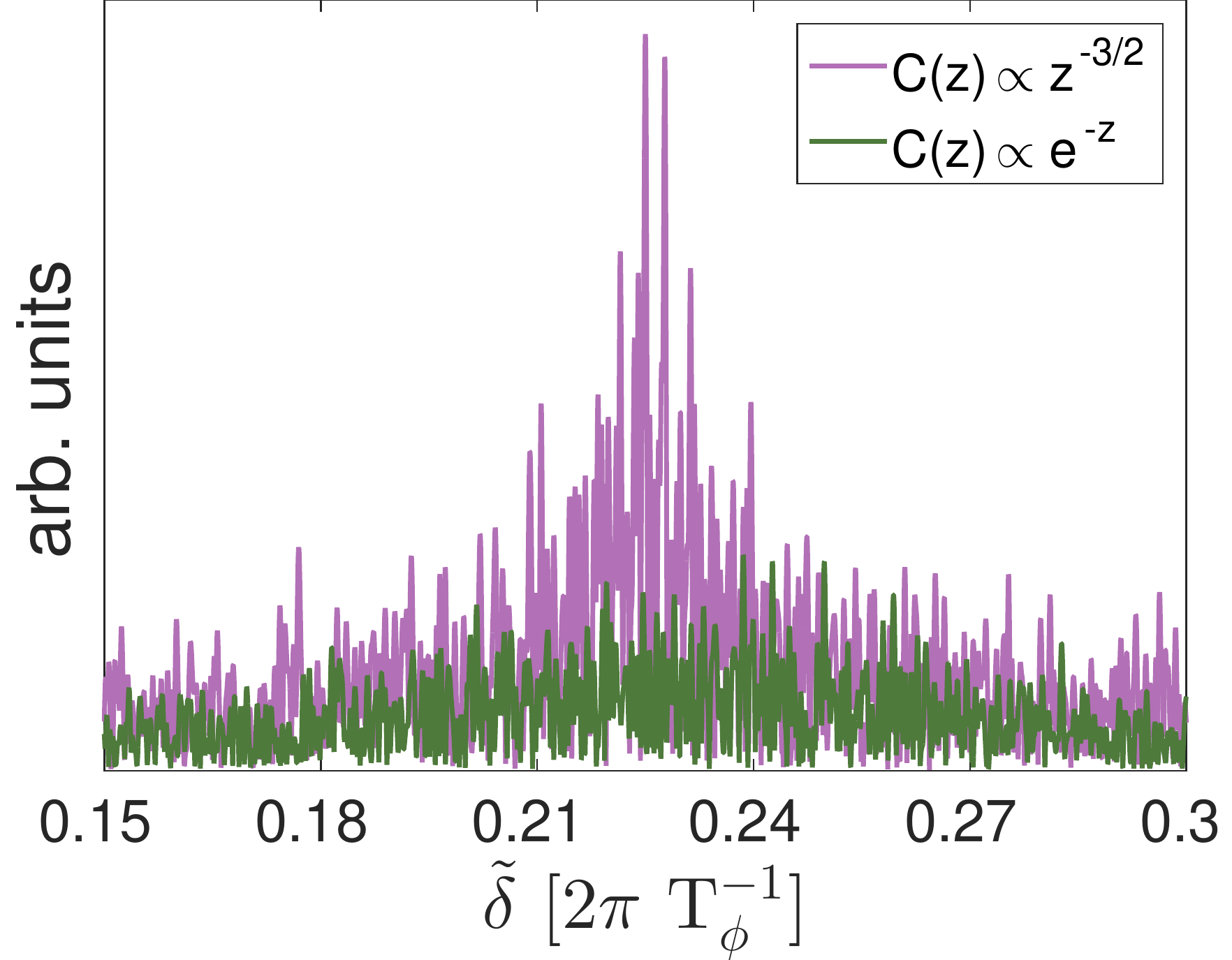}}
    \subfloat[\label{FigCFTfreqs:b}]{\includegraphics[width=5.8cm, height=4.455cm]{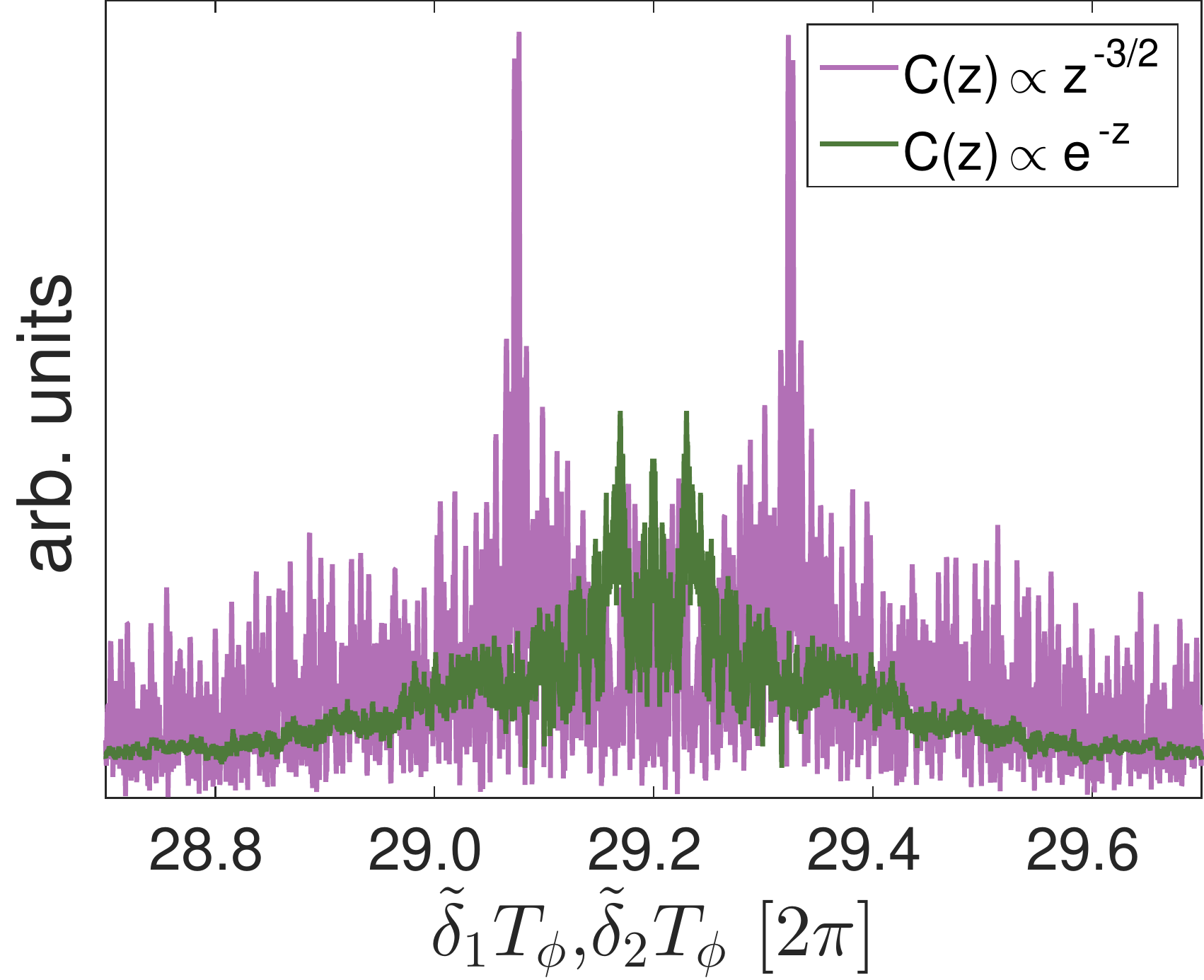}}
    \subfloat[\label{FigCFTfreqs:c}]{\includegraphics[width=5.8cm, height=4.455cm]{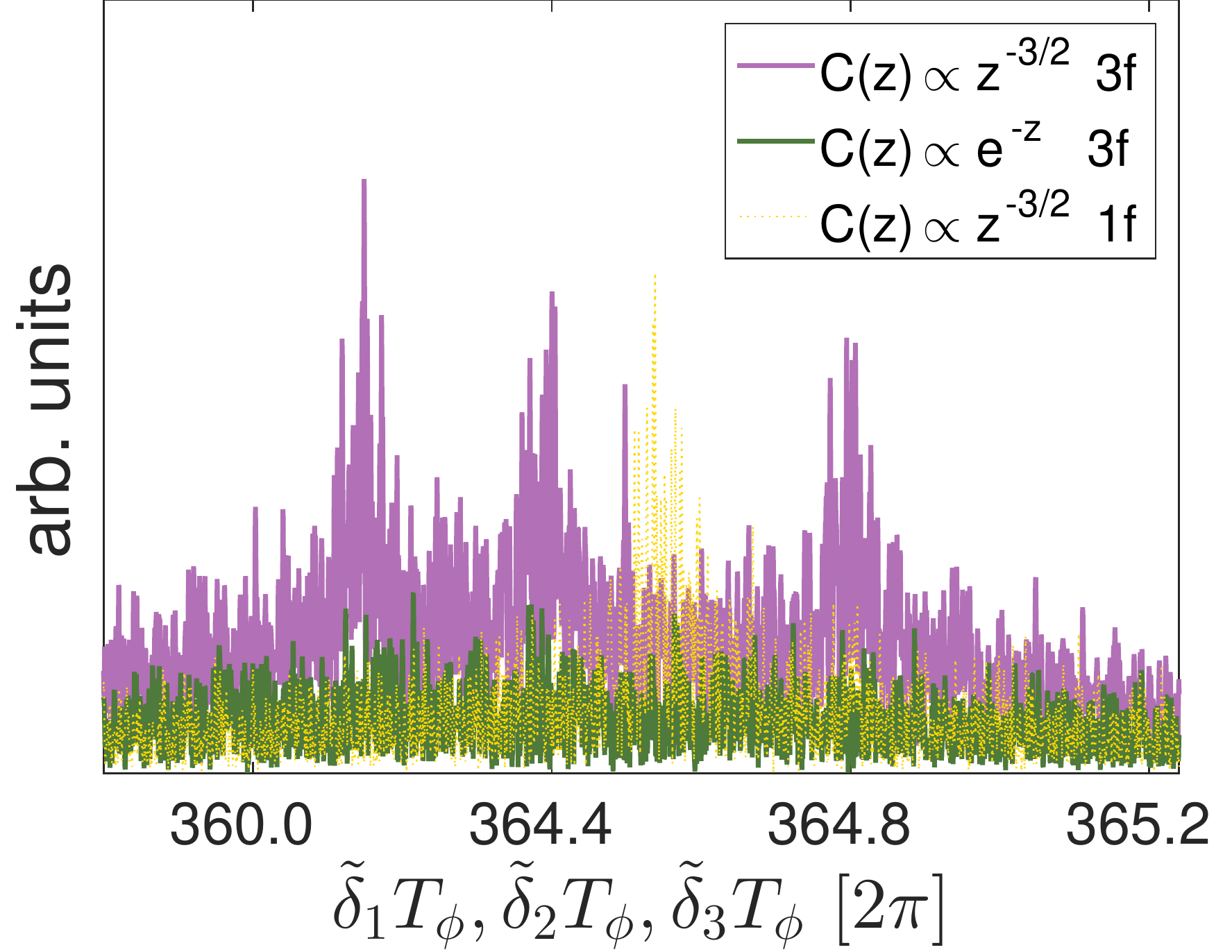}}
    \caption{Average of 400 CFT of measurement vectors for the cases of one frequency (a), two frequencies (b) and three frequencies (c), with the same parameters as used in the main text to generate the corresponding histograms. In purple, the case of $C(z^{-1.5})$ Eq. \ref{corr1} correlations whereas in green the correlations are exponential, which do not allow for frequency resolution.}\label{FigCFTfreqs}
\end{figure}

Correlation function fitting has the disadvantage that the noise produces displacements on the parameters, which manifest as a widening of histograms, but are less prone to local minima. In Fourier transform analysis, however, noise reflects appears as extra peaks that require more computation time to be avoided. Nonetheless, a direct Fourier transform of the signal can produce more visual results. 
In Fig. \ref{FigCFTfreqs} we present results for Continuous Fourier Transform (CFT) for the cases of signals containing one (a) two (b) and three (c) frequencies. Each CFT is calculated as 
\begin{equation}
CFT(\omega_i) = \sum_j s_j e^{-i\omega_i t}.
\end{equation}
Each plot is the average of 400 measurement vectors, where extended correlations allow for frequency resolution while exponential correlations produce a spectrum without defined peaks.

\section{Fisher Information \label{FI defnition}}
The Fisher Information (FI) matrix (for parameters \(\theta_i,\theta_j\)) is defined as 
 \begin{equation}
 	J_{i,j} = \mathop{{}\mathbb{E}}_{L(\theta)} \left[ \frac{\partial \log (L(\theta))}{\partial \theta_i} \frac{\partial \log ( L(\theta) )}{\partial \theta _j} \right],
 \end{equation}
 where \(L\) is the probability function, and \(\theta\) is a set of parameters that defines the probability.

\end{document}